\documentclass[]{spie}  

\usepackage{amsmath,amsfonts,amssymb}
\usepackage{graphicx}
\usepackage[colorlinks=true, allcolors=blue]{hyperref}

\usepackage{wrapfig}
\usepackage{sidecap}

\title{Polarization Modeling and Predictions for DKIST Part 1: Telescope and example instrument configurations.}

\author[a]{David M. Harrington}
\author[b]{Stacey R. Sueoka}
\affil[a]{National Solar Observatory, 8 Kiopa'a Street, Ste 201 Pukalani, HI 96768, USA}
\affil[b]{National Solar Observatory, 3665 Discovery Drive, Boulder, CO, 80303, USA}

\authorinfo{Further author information: (Send correspondence to D.M.H)\\D.M.H.: E-mail: harrington@nso.edu, Telephone: 1 808 572 6888\\  S.R.S.: E-mail: sueoka@nso.edu}

\begin{document} 
\maketitle

\begin{abstract}
We outline polarization performance calculations and predictions for the Daniel K. Inouye Solar Telescope (DKIST) optics and show Mueller matrices for two of the first light instruments. Telescope polarization is due to polarization dependent mirror reflectivity and rotations between groups of mirrors as the telescope moves in altitude and azimuth. The Zemax optical modeling software has polarization ray-trace capabilities and predicts system performance given a coating prescription. We develop a model coating formula that approximates measured witness sample polarization properties. Estimates show the DKIST telescope Mueller matrix as functions of wavelength, azimuth, elevation, and field angle for the Cryogenic Near Infra-Red Spectro-Polarimeter (CryoNIRSP) and for the Visible SpectroPolarimeter (ViSP). Footprint variation is substantial. We estimate 2\% variation of some Mueller matrix elements over the 5 arc minute CryoNIRSP field. We validate the Zemax model by show limiting cases for flat mirrors in collimated and powered designs that compare well with theoretical approximations and are testable with lab ellipsometers.
\end{abstract}

\keywords{Instrumentation, Polarization, Mueller matrix, DKIST, CryoNIRSP, ViSP}

\section{PREDICTING DKIST POLARIZATION}
\label{sec:intro}  

Predicting the Mueller matrix of a many-mirror system with highly powered optics across the field of view is an important tool for the design and use of large astronomical telescopes.   The Daniel K. Inouye Solar Telescope (DKIST) on Haleakal\={a}, Maui, Hawai'i has a 4.2m off-axis f/2 primary mirror (4.0m illuminated) and a suite of polarimetric instrumentation in a coud\'{e} laboratory \cite{2014SPIE.9145E..25M, Keil:2011wj, Rimmele:2004ew}. The telescope uses 7 mirrors to feed light to the coud\'{e} lab \cite{Marino:2016ks, McMullin:2016hm,Johnson:2016he,2014SPIE.9147E..0FE, 2014SPIE.9147E..07E, 2014SPIE.9145E..25M}. Operations involve 4 polarimetric instruments presently spanning the 380nm to 5000nm wavelength range.  A train of dichroic beam splitters allows for rapid changing of instrument configurations and simultaneous operation of 3 polarimetric instruments covering 380nm to 1800nm \cite{2014SPIE.9147E..0FE, 2014SPIE.9147E..07E, 2014SPIE.9147E..0ES, SocasNavarro:2005bq}. Complex modulation and calibration strategies are required for such a mulit-instrument system \cite{2014SPIE.9147E..0FE,2014SPIE.9147E..07E, Sueoka:2014cm, 2015SPIE.9369E..0NS, deWijn:2012dd, 2010SPIE.7735E..4AD}.  The planned 4m European Solar Telescope (EST), though on-axis, will also require similar calibration considerations \cite{SanchezCapuchino:2010gy, Bettonvil:2011wj,Bettonvil:2010cj,Collados:2010bh}.  Many solar and night-time telescopes are calibrating complex optical pathways \cite{DeJuanOvelar:2014bq, Joos:2008dg, Keller:2009vj,Keller:2010ig, Keller:2003bo, Rodenhuis:2012du, Roelfsema:2010ca, 1994A&A...292..713S, 1992A&A...260..543S,  1991SoPh..134....1A, Schmidt:2003tz, Snik:2012jw, Snik:2010ig,Snik:2008fh, Snik:2006iw, SocasNavarro:2011gn, SocasNavarro:2005jl, SocasNavarro:2005gv, Spano:2004ge, Strassmeier:2008ho, Strassmeier:2003gt, Tinbergen:2007fd, 2005A&A...443.1047B, 2005A&A...437.1159B}

Several other large astronomical telescopes are in development and include plans for polarimeters.  For many years, a night-time spectropolarimeter on the 4m Advanced Electro-Optical System (AEOS) telescope on Maui has been pursuing a campaign of polarization calibration \cite{Harrington:2015dl, Harrington:2010km, 2008PASP..120...89H, Harrington:2006hu, 2014SPIE.9147E..7CH}. We have developed Zemax modeling tools to compute the polarization of an optical system provided the optical model and the coating prescription for the optics.  These Zemax modeling tools have been used on the AEOS telescope and the HiVIS spectropolarimeter. We also apply the tools here to the DKIST telescope and a predict Mueller matrices for two of the first light polarimetric instruments. We refer the reader to recent papers outlining the various capabilities of the first-light instruments \cite{2014SPIE.9147E..07E, 2014SPIE.9145E..25M, 2014SPIE.9147E..0FE, Rimmele:2004ew}. 

In this work, we follow standard notation.  The Stokes vector is denoted as {\bf S} = $[I,Q,U,V]^T$. The Mueller matrix is the 4x4 matrix that transfers Stokes vectors. Each element of the Mueller matrix is denoted as the transfer coefficient \cite{1992plfa.book.....C, Chipman:2010tn, 2013pss2.book..175S}. For instance the coefficient (1,0) in the first row transfers $Q$ to $I$ and is denoted $QI$. The first row is denoted $II$, $QI$, $UI$, $VI$. The first column of the Mueller matrix is thus $II$, $IQ$, $IU$, $IV$.   

\subsection{DKIST Optics Overview}

The DKIST optical train includes an off-axis 4m diameter parabolic primary mirror (M1) that creates an f/2 prime focus.  At prime focus, there is a heat stop, which limits the field of view to roughly 5 arc minutes and reduces the heat load on all downstream optics.  The secondary mirror (M2) is also an off-axis ellipse (conic -0.54), which relays this beam to an f/13 Gregorian focus. Just above the Gregorian focus, there is an optical station for insertion and removal of several masks, targets, an artificial light sources and a set of polarization calibration optics. There are also field stops for 2.8 arc minute diameter and 5 arc minutes diameter at Gregorian focus. The third mirror (M3) is a flat fold mirror at 45$^\circ$ incidence angle that directs the light towards the off axis ellipse (M4, conic -0.37). This parabolic mirror changes the diverging f/13 beam to a converging f/53 beam and creates a pupil conjugate plane near the next flat steering mirror (M5). The elevation axis for the telescope is also between M4 and M5. M5 folds at 30$^\circ$ and also functions as a fast steering mirror with tip/tilt capability. The sixth mirror in the system (M6) is also a flat and directs the beam vertically downward toward the coud\'{e} lab folding at a 60$^\circ$ angle.  The seventh mirror is another flat fold mirror that levels the beam into the coud\'{e} laboratory folding at 90$^\circ$ incidence angle.  The eighth mirror (M8) is an off axis parabola that collimates the beam.  The ninth mirror is a flat fold that directs the beam towards the deformable mirror as part of the integrated adaptive optics system.  The pupil of the system is conjugated near the flat deformable mirror (DM) and this represents the 10th optic in the system (M10).    

\begin{figure} [htbp]
\begin{center}
\begin{tabular}{c} 
\hbox{
\hspace{-1.5em}
\includegraphics[height=5.5cm, angle=0]{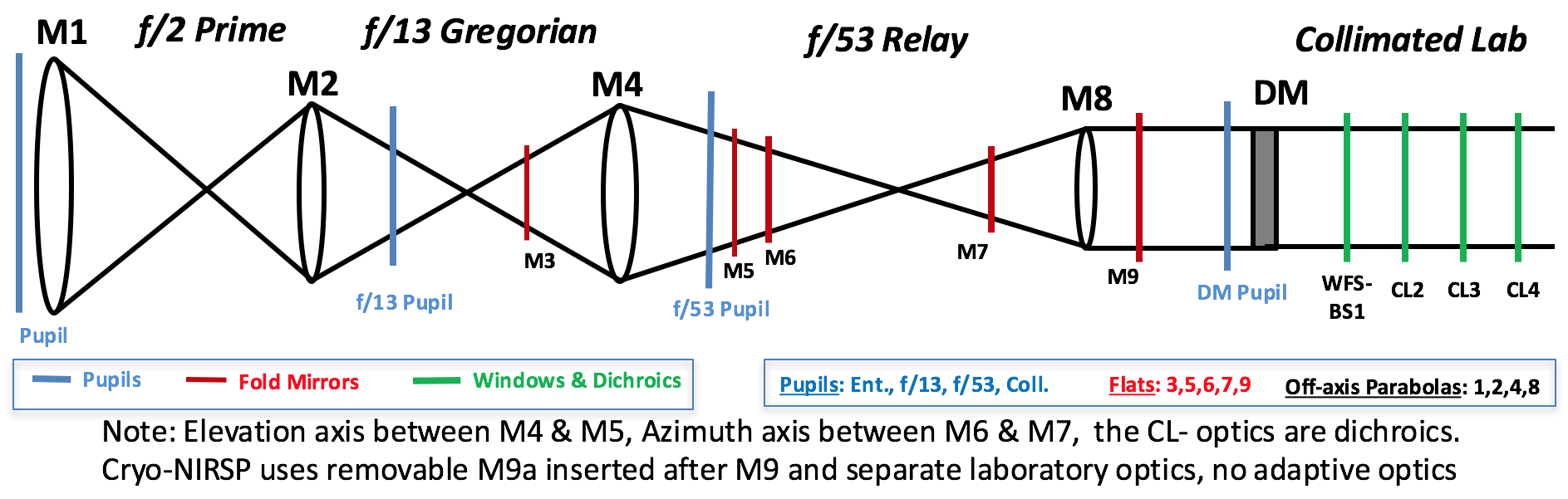}
}
\end{tabular}
\end{center}
\caption[optical_concept] 
{ \label{fig:optical_concept} 
The DKIST telescope feed optics conceptual design. Only powered optics are shown.  All powered optics are off-axis parabolas.  The beam has an f/2 prime focus, an f/13 Gregorian focus, an f/53 intermediate focus and a collimated coud\'{e} laboratory where all first-light instruments are installed.  All flat mirrors are shown as red lines.  Pupil conjugate planes are shown as vertical blue lines. The entrance pupil of this unobscured system is a 4.0m diameter aperture mask ring mounted above the 4.24m diameter primary mirror.  The deformable mirror (DM) is shown as a thick grey box.  After the deformable mirror, there are several beam splitters as part of the coud\'{e} laboratory.  The wave front sensor of the adaptive optics system is fed by a reflection off the uncoated front surface of a window, denoted WFS-BS1.  After this optic, there are several dichroic beam splitters that are interchangeable denoted as CL{\it N} where {\it N} is the numeric identifier.  These dichroics are designed to be reconfigured to send various wavelength ranges to various instruments.  We also note that one of the polarimetric instruments (Cryo-NIRSP) inserts a mirror after M9 and does not use the adaptive optics system.  }
\end{figure}

Figure \ref{fig:optical_concept} shows the optical concept for the system.  There are 4 powered optics that perform the relays.  In order, the beam has an f/2 prime focus, an f/13 Gregorian focus, an f/53 intermediate focus and a collimated coud\'{e} laboratory.  

The coud\'{e} lab was designed to allow simultaneous operation of many instruments observing different wavelengths by using several custom dichroic beam splitters.  An adaptive optics system was also integrated into the design. The wave front sensor (WFS) of the adaptive optics system is fed by a reflection off the uncoated front surface of a window, denoted WFS-BS1 that is mounted after the DM.  There are 3 polarimetric instruments that use the adaptive optics system: the Visible Spectro-Polarimeter  (ViSP), the Visible Tunable Filter (VTF) and the Diffraction Limited Near-InfraRed Spectro-Polarimeter  (DL-NIRSP).  In addition to these polarimetric instruments, there are two arms of a high resolution imaging system that also use the AO feed.  The Visible Broadband Imager (VBI) is essentially two separate instruments, one a red imager (VBI-red) and the other a blue imager (VBI-blue).  

Another first light instrument was designed to include infrared capabilities at wavelengths as long as 5000nm and was optimized for seeing-limited science cases.  This instrument, the Cryogenic Near-Infrared Spectro-Polarimeter (CryoNIRSP) does not use the adaptive optics system and has a separate optical path after M9.  For CryoNIRSP, an additional fold mirror (M9a) is inserted into the beam after M9 to direct light to the system feed optics. 

\begin{figure} [htbp]
\begin{center}
\begin{tabular}{c} 
\hbox{
\hspace{-1.5em}
\includegraphics[height=8.6cm, angle=0]{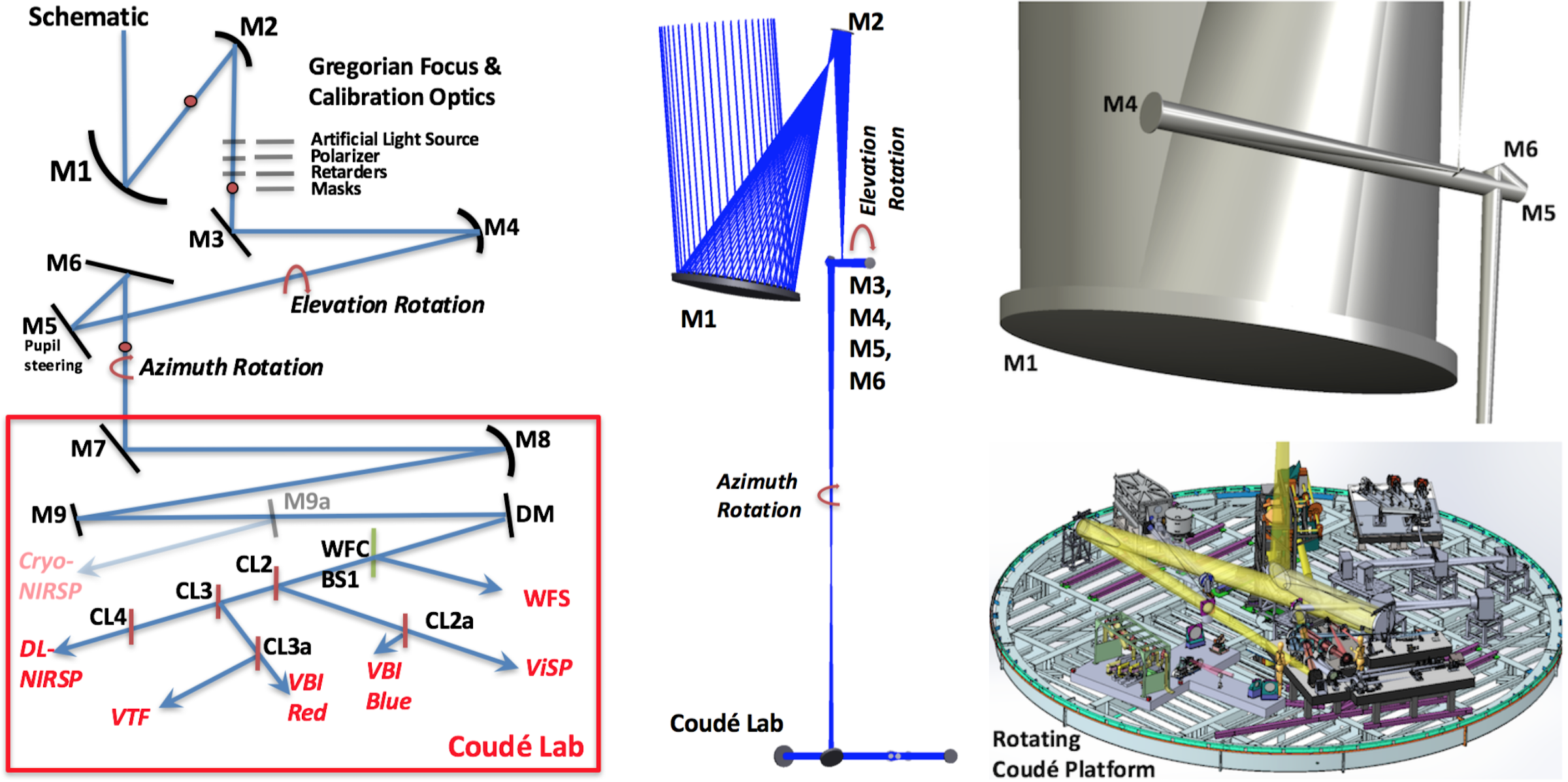}
}
\end{tabular}
\end{center}
\caption[zemax_layout] 
{ \label{fig:optical_outline} 
The DKIST optics design and schematic. The left panel shows a cartoon schematic for the optical path (not to scale, not an actual layout, fold angles not accurate). The primary and secondary mirrors relay the beam to the f/13 Gregorian focus and an optical system where calibration optics can be inserted. The M3 through M6 optics assist with the articulation of the telescope, and feed the f/53 beam down to the coud\'{e} laboratory. The beam is then leveled and collimated in to the adaptive optics system by M7 through M10.  The optional insertion of a flat mirror to bypass the adaptive optics and feed the CryoNIRSP instrument is also shown. The wave-front sensor beam splitter (WFS-BS1) and all the dichroic beam splitters (CL{\it N}) are shown distributing the different wavelengths to all the coud\'{e} instruments.  The middle panel shows the Zemax design beginning from the beam entering the dome and ending at the deformable mirror in the coud\'{e} laboratory. Mirrors M3 through M6 are packaged in an assembly that allows the telescope structure to rotate in elevation.  The top right image shows a solid model of the optical beam highlighting the Gregorian focus and the optics M3 through M6.  The primary (M1) and the 4m diameter beam is seen behind optics M3-M6.  The bottom right image shows a recent solid model of the coud\'{e} laboratory.  The beam comes to the lab from above M7 and is highlighted in yellow.  M7 folds the beam at a 90$^\circ$ angle to be level with the lab floor.  The entire laboratory is a rotating platform and is another degree of freedom for the optical system.  Not all beams in all instruments are shown in the lower right hand solid model. }
\end{figure}

For calibration purposes, we will describe a possible configuration using the AO system and the several simultaneous channels of the various instruments. Note that both ViSP and DL-NIRSP have three separate cameras that can record three separately configured wavelengths each. Many configurations are possible and the spectrographs are designed to be re-configurable in minutes with substantial automation.  As an example, DL-NIRSP configuration of the three spectrographs could be (789 nm or 854.2 nm) on camera 1,  (1074.7nm or 1083 nm) on camera 2,  (1430nm or 1565nm) on camera 3.  At the same time, ViSP could be configured to a vast array of possible spectral lines covering 380nm to 1100nm.  Depending on how the dichroics are arranged, the various instruments could be sent limited ranges of wavelengths permitting only some of the cameras to be used.  

One setup could configure the first two dichroics CL2 and CL2a to send the VBI-blue camera wavelengths shorter than 430nm, ViSP wavelengths to 660nm, VTF to 860nm and DL-NIRSP the long wavelength bandpass.  With this setup, ViSP could be configured to use at least two of the cameras between 430nm and 660nm, and the DL-NIRSP could use two of the three cameras working at wavelengths longer than 1000nm.  We would need to have polarization calibrations for 5 polarimetric channels (2 on ViSP, VTF, 2 on DL-NIRSP) with calibrations done after the dichroics are installed.  

\begin{figure} [htbp]
\begin{center}
\begin{tabular}{c} 
\hbox{
\hspace{-1.5em}
\includegraphics[height=11.7cm]{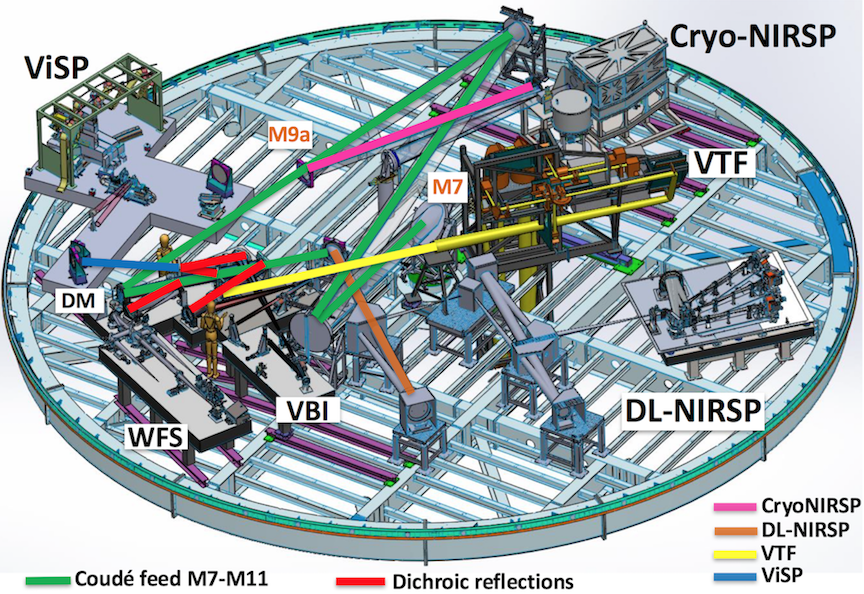}
}
\end{tabular}
\end{center}
\caption[articulated] 
{ \label{fig:coude_lab_annotated} 
The DKIST coud\'{e} laboratory and the various optical pathways feeding the instruments. The green line shows the coud\'{e} beam beginning from the reflection off the M7 flat. The M8 parabola collimates the beam and the beam is then folded by M9. There is a position labeled M9a where the optional CryoNIRSP pickoff flat can be inserted to feed all the light to CryoNIRSP following the magenta line.  For all other instruments, the beam continues past the M9a station following the green line to the DM and is then AO corrected.  There are several short red lines in a very confined space representing all the dichroic beam splitters.  Ultimately these many reflections send different wavelengths to the polarimetric instruments (ViSP, VTF and DL-NIRSP) along with the various camera and sensor systems (the adaptive optics WFS and the two high resolution imaging cameras VBI-red, VBI-blue).  The blue line shows the light path leaving the dichroic feeding ViSP. The rest of the ViSP beam is not shown on the ViSP optical bench. Yellow shows the light path leaving another dichroic feeding VTF.  Orange shows a beam transmitting through the last dichroic and being reflected into DL-NIRSP.  }
\end{figure} 

Figure \ref{fig:optical_outline} shows a conceptual optical schematic along with Zemax and solid models of the system.  The telescope is a classical elevation over azimuth design.  In addition to the usual azimuth and elevation degrees of freedom, the entire laboratory floor rotates freely.  This coud\'{e} rotator is one of the major construction components, and gives independent optical control of the field rotation without requiring the use of a 3-mirror derotator (K cell) \cite{2014SPIE.9145E..25M} or more optics such as THEMIS \cite{LopezAriste:2011wc} or the EST \cite{Bettonvil:2011wj,Bettonvil:2010cj}.   Figure \ref{fig:coude_lab_annotated} shows the optical beam on the lab floor, feeding all instruments.  The optional CryoNIRSP feed is shown as a semi-transparent line.  All other instruments can be operated simultaneously with the adaptive optics.  

In the optical modeling efforts, all three degrees of freedom (Azimuth, Elevation, Coud\'{e} table angle) impact polarization calibration plans through the rotation of the projected image against the solar disk and subsequent rotation of polarization calibrations when tracking with images fixed to parallactic or solar coordinates.  The azimuth and coud\'{e} table angles are redundant optically, but do repoint the system celestially. We must consider the relative image rotation angle and time dependent Mueller matrices when tracking the sun under a variety of use cases that either use or do not use the table angle.

\begin{wraptable}{r}{0.6\textwidth}
\vspace{-3mm}
\caption{Optical Properties of the DKIST On-axis Beam with AOI}
\label{table:AOI}
\centering
\begin{tabular}{c c c c c c c c}
\hline\hline
Optic		& AOI 	& AOI  	& AOI 	& AOI 	& f/\# 	& f/\# 	& Coating	\\
Name	& Chief 	& Max 	& Min	& Range	& In 		& Out	& 		\\
\hline
M1		& 14.04	& 20.56 	& 7.13 	& 13.4 	& $\infty$ 	& 2		& Al-Al$_2$O$_3$ 	\\
M2		& 11.84 	& 17.27 	& 6.03	& 11.3 	& 2 		& 13 		& Enh.Ag 		\\
M3 		& 45		& 47.19 	& 42.81 	& 4.4 	& 13 		& 13 		& Enh.Ag 		\\
M4 		& 1.76 	& 2.57 	& 0.93 	& 1.6 	& 13 		& 53		& Enh.Ag 		\\
M5 		& 15 		& 15.53 	& 14.47 	& 1.1 	& 53 		& 53 		& Enh.Ag 		\\
M6 		& 30 		& 30.53 	& 29.47 	& 1.1 	& 53 		& 53 		& Enh.Ag 		\\
\hline
M7 		& 45 		& 45.53 	& 44.47 	& 1.1 	& 53 		& 53 		& Enh.Ag 		\\
M8 		& 5.33 	& 5.60 	& 5.06 	& 0.5 	& 53 		& $\infty$ 	& Enh.Ag 		\\
M9 		& 10 		& 10		& 10  	& 0 		& $\infty$ 	& $\infty$ 	& Enh.Ag 		\\
DM 		& 15 		& 15 		& 15 		& 0 		& $\infty$ 	& $\infty$ 	& Enh.Ag 		\\
BS1 		& 15 		& 15 		& 15 		& 0 		& $\infty$ 	& $\infty$ 	& None 		\\
CL2	 	& 15 		& 15 		& 15 		& 0 		& $\infty$ 	& $\infty$ 	& Dichroic		\\
{\it CL2a} 	& 15 		& 15 		& 15 		& 0 		& $\infty$ 	& $\infty$ 	& Dichroic		\\
CL3		& 15 		& 15 		& 15 		& 0 		& $\infty$ 	& $\infty$ 	& Dichroic		\\
{\it CL3a}	& 15 		& 15 		& 15 		& 0 		& $\infty$ 	& $\infty$ 	& Dichroic		\\
CL4		& 15 		& 15 		& 15 		& 0 		& $\infty$ 	& $\infty$ 	& Dichroic		\\
\hline
\end{tabular}
The angle of incidence (AOI) for the on axis (zero field) beam. The chief ray is traced through the system as well as the lowest and highest AOI for the marginal rays for the on axis beam (zero field of view) on the optics.  For instance, the primary mirror (M1) reflects the chief ray at 14.04$^\circ$. Since M1 is a tilted off-axis parabola, the marginal rays are incident at angles between 7.13$^\circ$ and 20.56$^\circ$ representing a range of 13.4$^\circ$ AOI.  The beam sees smaller AOI variation across the beam footprint in the f/53 portion of the relay.  In the collimated beam after M8, there is no AOI variation across a footprint (zero field). The various dichroics combine in reflection and / or transmission to feed the beam to various instruments. See Figures \ref{fig:optical_outline} and \ref{fig:coude_lab_annotated} as well as the text.
\vspace{-4mm}
\end{wraptable}

When considering polarization performance of the optical system, the angle of incidence (AOI) variation as well as the variation across the field of view (FoV) are both important considerations. As an example, the DKIST primary mirror converts a collimated beam from a single field  angle to an f/2 converging beam. The effective fold angle for M1 is roughly 28.1$^\circ$ , but the bundle of rays exiting the optic sees fold angles between 14.3$^\circ$ and 41.1$^\circ$ across the beam footprint. The polarization properties vary strongly with AOI and this imparts polarization variation as a function of position in the beam leaving M1.  Table \ref{table:AOI} shows the variation in incidence angle for the on axis (zero-field) beam in the design. The first column {\it Optic Name} lists the optic. The next 4 columns show the AOI for the chief ray of the beam, the the marginal ray with maximum and minimum AOI and the range of incidence angles.  Subsequent columns of Table \ref{table:AOI} show the effective f/ number of the incoming and outgoing beams as well as the coating on the optic. The primary mirror is coated with bare aluminum, which quickly forms a thin oxide layer. All other mirrors M2 to the DM are coated with enhanced protected silver down to the coud\'{e} lab.  These multi-layer coatings tend to have stronger dependence on polarization properties with incidence angle and hence are important to model accurately across the full field of view. The wave front sensor beam splitter (WFS-BS1) has an uncoated front surface to feed the Fresnel reflection to the adaptive optics WFS.  The coud\'{e} lab dichroic beam splitters CL{\it N} all have custom coatings to reflect some wavelengths while efficiently transmitting all other wavelengths.  All beam splitters have an anti-reflection coating on the back surface optimized for their specific transmission wavelength region. 

Whenever the pupil is demagnified, the field variation of incidence angles is increased accordingly. The incidence angle variation on the primary mirror is also the field of view angle. The optics demagnify the entrance pupil onto the deformable mirror, which is only 0.2m across giving a 20x demagnification from the 4m entrance aperture. We trace the incidence angle variation in Table \ref{table:FoV} for the chief ray for every field point in the 2.8 arc minute diameter field. As powered optics change the relationship between angles across the beam, we list the input and output variation for each powered optic where changes occur. The {\it Beam Loc} column in Table \ref{table:FoV} shows the surface where AOI variation with FoV is computed.  The next four columns show the field center incidence angle and the Min/Max incidence angles at the FoV edge. The field of view of the primary mirror is the nominal 2.8 arc minutes (0.05$^\circ$). However, as the optics demagnify the beam, the incidence angles increase across the field. In the coud\'{e} lab, the pupil on the DM sees an incidence angle variation with FoV of 0.9$^\circ$, roughly 20x the original 2.8 arc minute FoV.  It's apparent from Table \ref{table:FoV} to see that the field variation is roughly one degree in all optics of the f/53 beam and in the collimated beam, which is the dominant source of field variation effects for polarization calibration.

\section{Zemax Computations}

Zemax traces individual rays in the Jones formalism through a geometric model.  Zemax can propagate rays from any position in the entrance pupil at any field angle through the optical design. In Zemax, we have developed a script to trace polarized rays across the pupil and field while specifying a series of wavelengths, polarization states and system optical configurations. We have adapted scripts initially developed by Don Mickey in 2002 \cite{Harrington:2006hu}. 

\begin{wraptable}{l}{0.5\textwidth}
\vspace{-0mm}
\caption{Properties of the chief ray with FoV}
\label{table:FoV}
\centering
\begin{tabular}{c c c c c c}
\hline\hline
Beam	& AOI 	& AOI  	& AOI 	& AOI 	& Beam	 		\\
Loc.		& (0,0) 	& Max 	& Min	& w/ FoV	&  f/\# 	 		\\
\hline
M1 In	& 14.00	& 14.02 	& 13.98 	& 0.05 	& $\infty$ 	  		\\
M2 In	& 11.84	& 11.93 	& 11.76 	& 0.17 	& 2 		  		\\
M2 Out	& 4.39 	& 4.57 	& 4.20	& 0.38 	& 13 		  	  	\\
M3 		& 45		& 45.19  	& 44.81 	& 0.38 	& 13 		  	  	\\
M4 In	& 1.76 	& 1.89 	& 1.63 	& 0.38 	& 13 		 	  	\\
M4 Out	& 3.52 	& 3.98 	& 3.07 	& 0.90 	& 53 		 	  	\\
M5 		& 15 		& 15.38 	& 14.63 	& 0.90 	& 53 		  	  	\\
M6 		& 30 		& 30.38	& 29.63 	& 0.90 	& 53 		  	  	\\
\hline
M7 		& 45 		& 45.45 	& 44.56 	& 0.90 	& 53 		   		\\
M8 Out	& 10.66 	& 11.11 	& 10.22 	& 0.90 	& $\infty$ 		    	\\
M9 		& 10 		& 10.45	& 9.54 	& 0.90 	& $\infty$ 	    		\\
DM 		& 15 		& 15.45	& 14.55	& 0.90 	& $\infty$ 	    		\\
BS1 		& 15 		& 15.45 	& 14.55	& 0.90	& $\infty$ 	    		\\
CL2 		& 15 		& 15.45 	& 14.55	& 0.90	& $\infty$ 	    		\\
\hline
\end{tabular}
The angle of incidence (AOI) variation with field of view (FoV). The table lists the incidence angle for the chief ray for every field point for the 2.8 arc minute field propagating through the coud\'{e} lab and adaptive optics system.  See text for details.
\end{wraptable}

\begin{wrapfigure}{r}{0.45\textwidth}
\vspace{-100mm}
\begin{center}
\begin{tabular}{c} 
\hbox{
\hspace{-1.5em}
\includegraphics[height=6.6cm]{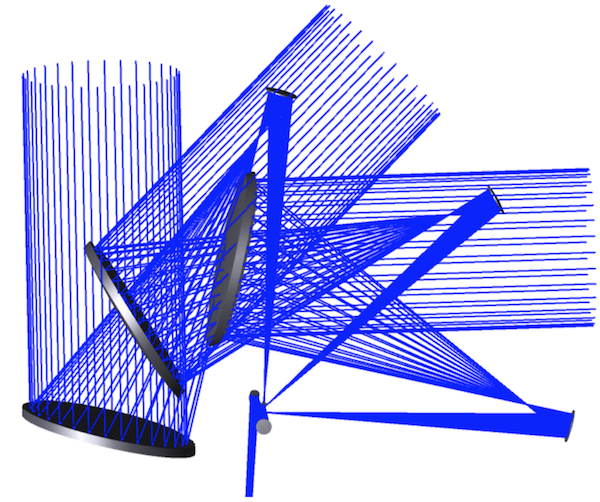}
}
\end{tabular}
\end{center}
\caption[articulated] 
{ \label{fig:articulated} 
The DKIST telescope design articulated in elevation. The off-axis f/2 primary mirror is seen in multiple locations as the telescope assembly pivots about the elevation axis.}
\end{wrapfigure}

The new scripts can change fold angles, rotation angles, wavelengths, etc, in order to provide the ability to simulate a wide range of optical system configurations.  With this functionality, we can derive polarization properties of the system across the beam footprint for any desired setting on any desired surface in the system. We have recently applied this analysis to the 4m AEOS telescope and compared the predictions to polarization calibrations derived from the daytime sky \cite{Harrington:2017ej}. 

Depending on the sensitivity and computational speed required, the pupil sampling, field sampling, wavelength coverage, and telescope pointing step size can be increased to sample the beam to a desired accuracy.  Typically sampling the footprint in a 20x20 grid of rays achieves $<$0.0001 level numerical precision, consistent with our simulation needs and is a good compromise between computation speed and numerical accuracy.  For other systems, more fine sampling may be desired when confronting more complex situations such as assessing vignetting at each field angle and in general the symmetries in the polarization properties of the final beam footprint when polarization analysis is performed.  

The Zemax scripts output over 30 electric field vector components for every ray traced. The optical surface can also be specified to examine polarization properties on any optic in the design. To compute a Muller matrix from the Jones formalism, we independently trace a set of the six purely polarized inputs $\pm Q$, $\pm U$ and $\pm V$ through the system.  For each of the six inputs, Zemax calculates the electric field properties in 3-D at the specified surface.

\begin{figure*}[htbp]
\begin{center}
\includegraphics[width=0.99\linewidth, angle=0]{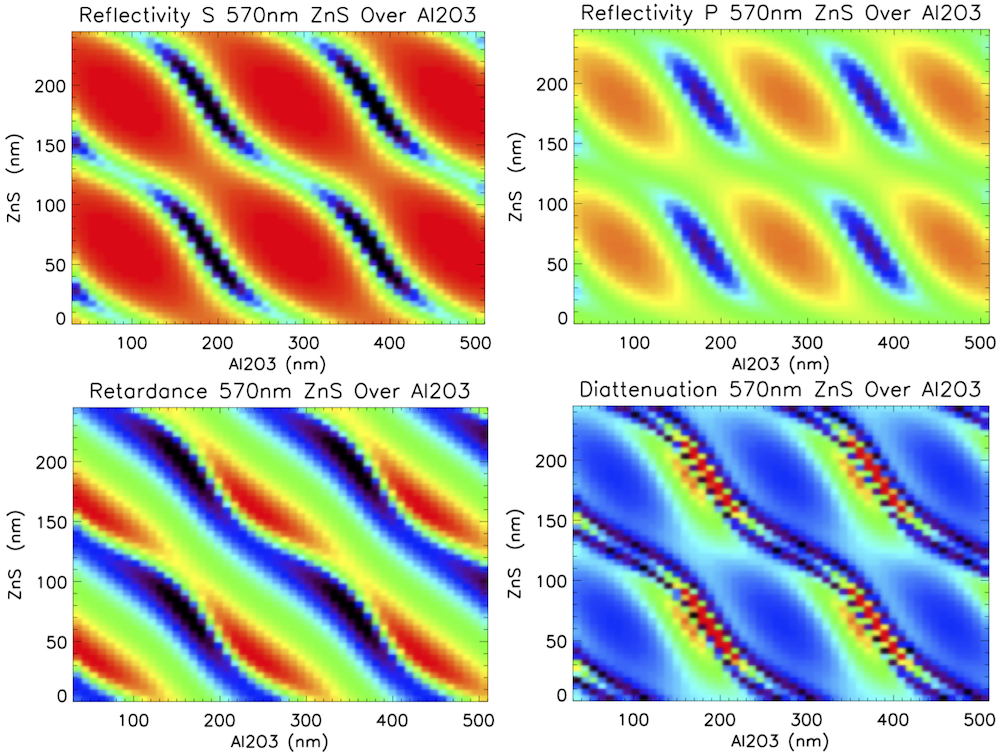}
\caption{The Zemax coating calculations as layer thickness is varied at 570nm wavelength and 45$^\circ$ incidence angle. We ran a 2-layer overcoating of zinc sulfide (ZnS) coated over aluminum oxide (Al2O3) coated over the silver base coating. For each panel we ran a grid of 50 by 50 thicknesses. The x-axis for every panel shows varying thickness of the aluminum oxide from 0 to 500nm.  The y axis for each panel shows the thickness of ZnS from 0 to 250nm. Note that for completeness, we ran models at half and double these scales (not shown here) to verify behavior of thicker and thinner layers.  We also ran all combinations of ZnS, Al$_2$O$_3$, MgF$_2$ and SiO$_2$ (not shown here).  The top left panel shows the reflectivity for the S polarization state. The linear color scale runs from black at 85\% reflectivity to red at 100\% reflectivity.  The top right panel shows the reflectivity for the P polarization state on the same color scale (85\% to 100\%). The bottom left panel shows the retardance. The linear color scale runs from black at 140$^\circ$ to red at 215$^\circ$. The bottom right panel shows the diattenuation. The color scale goes from 0\% for black to 3\% for red. For all the plots, there are regions of high and low reflectivity, diattenuation and retardance corresponding to the coherent effects of the coating layer thicknesses.  }
\label{fig:coating_spaces}
\end{center}
\end{figure*}

These electric field calculations are turned in to Stokes vector formalism for each of the pure input states. The computed intensity is the square of the XY components of electric field amplitude (ExEx + EyEy).  Stokes $Q$ is the X and Y amplitude difference: (ExEx - EyEy).  The term $\delta$ represents the phase difference and is one of the Zemax outputs, but can also be computed directly as $\delta$ = $\phi_x$ - $\phi_y$. Stokes $U$ is computed from X and Y electric field amplitudes accounting for coherent phase variations: 2ExEyCOS($\delta$). Stokes $V$ is similarly computed with both XY field amplitudes and phases: 2ExEySIN($\delta$).

We demonstrate Zemax polarization properties in this paper by computing the electric field distributions, Stokes vectors across various footprints and Mueller matrices for various optical systems. Figure \ref{fig:articulated} shows the articulation of the DKIST optical design from the primary mirror to the 6th mirror in the system (M6), which represents the optical configurations determining the azimuth-elevation pointing of the system as traced by our scripts. In the coming sections we will show the Zemax computations for a simple flat fold mirror in a powered beam, and then a variety of surfaces within the DKIST design.

\begin{figure}[htbp]
\begin{center}
\hbox{
\hspace{-1.0em}
\includegraphics[width=0.99\linewidth, angle=0]{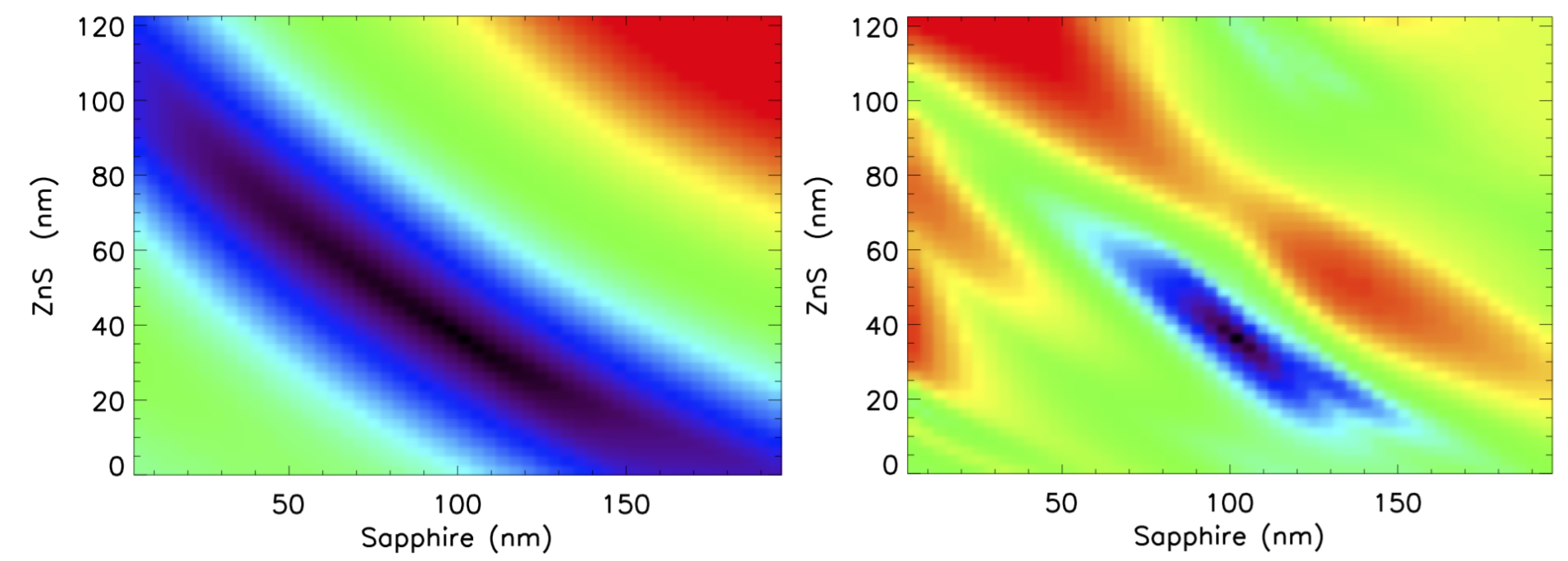}
}
\caption{The differences between a particular coating model and the various model coating formulas computed in Zemax for two example coating materials. The absolute value of the differences between retardances and diattenuations were summed over all wavelengths for every combination of material thicknesses. Low differences are color coded blue/black while large differences are red. The left panel shows the difference between retardance values.  The right panel shows the difference between diattenuation values.  Note, the behavior of diattenuation and retardance errors are quite different. We tested 3 common materials with one material coated over another for a total of 6 coating formulas.}
\label{fig:coating_search_space}
\end{center}
\vspace{-12mm}
\end{figure}

\section{Coating Formulations}

The complex refractive index and thickness of each layer in a dielectric coating impact the polarization performance.  For typical enhanced-protected metal coatings, there are one to several dielectric layers coated on top of the metal layer. Coatings are typically optimized for reflectivity over certain wavelength ranges but also be optimized for retardation and diattenuation. 

Multi-layer coatings can create two or more wavelengths where the retardation near the theoretical 180$^\circ$ for a perfect reflection. They can also introduce substantial retardation and diattenuation at other wavelengths, which depends strongly on incidence angles. The DKIST calibration plan presently groups the telescope feed optic Mueller matrices and reduces the number of variables required to predict the telescope Mueller matrix for all wavelengths, fields and telescope pointings \cite{2014SPIE.9147E..0FE, 2010SPIE.7735E..4EE, 2014SPIE.9147E..07E}. To create estimates of the likely DKIST Mueller matrix dependencies on field, telescope pointing and wavelength, we need a model for the coating formula that captures the relevant dependencies on incidence angle and wavelength.

DKIST internal studies reported measurements of the retardation and reflectivity for witness samples across the 400nm to 900nm wavelength range. To estimate DKIST polarization performance, we needed our model coating formula to be representative of the expected retardance, diattenuation and reflectivity. 

Without access to a manufacturer-provided formula, we found a simple search of standard coating materials identified a reasonable formula for the coating that matched reflectivity, diattenuation and retardance. For our polarization performance calculations, the retardance and diattenuation are important to match than the overall reflectivity.  With an approximate coating formula, we can to estimate the amplitude of several polarization effects expected in DKIST. Having a formula allows us to estimate the expected dependence on incidence angle or field of view with reasonable amplitudes \cite{Sueoka:2016vo}. We wrote a Zemax script to output a coating polarization report for many combinations of material thicknesses allowing an efficient search of several possible coating formulas. For enhanced and / or protected silver formulas, fused silica (SiO$_2$), zinc sulfide (ZnS), magnesium flouride (MgF$_2$) and aluminum oxide (Al$_2$O$_3$) can be used as the protected layer.  An example coating model run at 570nm wavelength is shown in Figure \ref{fig:coating_spaces}.  Several additional coating formulas are shown in our recent publication \cite{Harrington:2017ej}.

\begin{wrapfigure}{r}{0.6\textwidth}
\centering
\vspace{-0mm}
\hbox{
\hspace{-0.5em}
\includegraphics[width=0.56\textwidth]{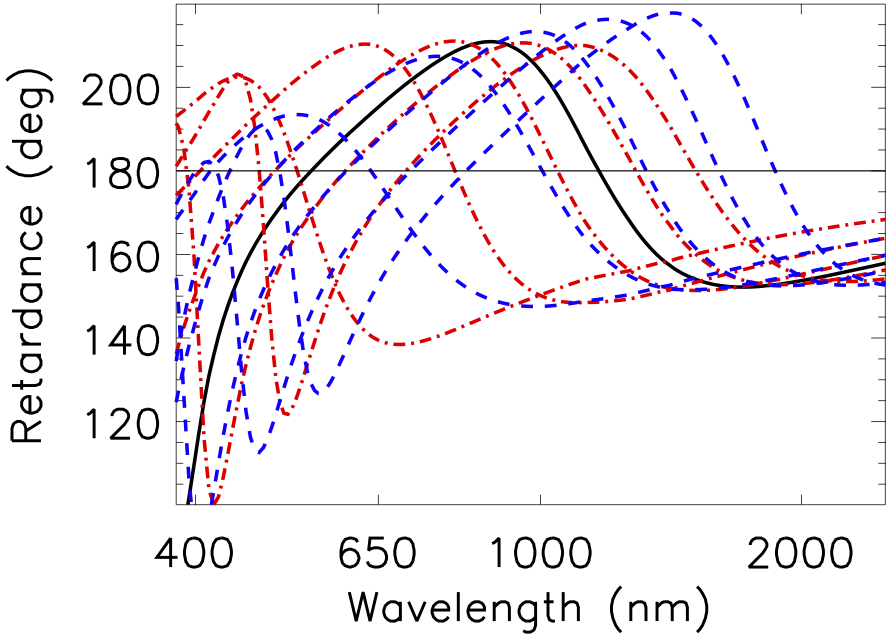}
\hspace{0mm}
}
\caption{The model retardance versus wavelength for selected coating layer thickness drawn from Figure \ref{fig:coating_search_space}.  There are strong wavelength dependent changes with even a few nm thickness variation.  Black shows the nominal curve of 37.5nm ZnS over 100nm of Al$_2$O$_3$.  The red curve shows variations in ZnS thickness over 100nm of  Al$_2$O$_3$.  The blue curves show variations in  Al$_2$O$_3$ with 37.5nm of ZnS overcoated. }
\label{fig:coating_variation}
\vspace{-4mm}
\end{wrapfigure}

Often the harder materials (sapphire, fused silica) are used as the durable protective over-coating while other layers or materials are included to minimize retardance or maximize reflectivity at particular wavelengths with thickness tolerances of a few nm. All of the 2-layer protective coating formulas shown in \cite{Harrington:2017ej} and the searches reported here do have two separate 180$^\circ$ retardance crossing wavelengths around 400nm and 850nm as was desired by the DKIST project.

We searched the common materials of ZnS, SiO$_2$, MgF$_2$ and Al$_2$O$_3$ in 2-layer protective coatings over the metal layer.  As an example of one of these searches, Figure \ref{fig:coating_search_space} shows a search of up to 200nm aluminum oxide over a layer of up to 120nm of zinc sulfide. Figure \ref{fig:coating_variation} shows variations in retardance for the reflected beam with 5nm changes in thickness of two dielectric layer thicknesses (ZnS and Al$_2$O$_3$). We identified a coating formula that has similar retardation, reflectivity and diattenuation to our witness samples for the DKIST mirrors. This coating formula was not an exhaustive search of possible design space, but simply a few iterations to achieve a reasonable match. For modeling efforts presented here, this coating formula will be useful to predict the system Mueller matrix for the CryoNIRSP instrument. We show the 1000nm to 5000nm wavelength range and select wavelengths for the ViSP instrument where the model coating formula matches the witness sample retardance.

\begin{figure}[htbp]
\begin{center}
\includegraphics[width=0.99\linewidth, angle=0]{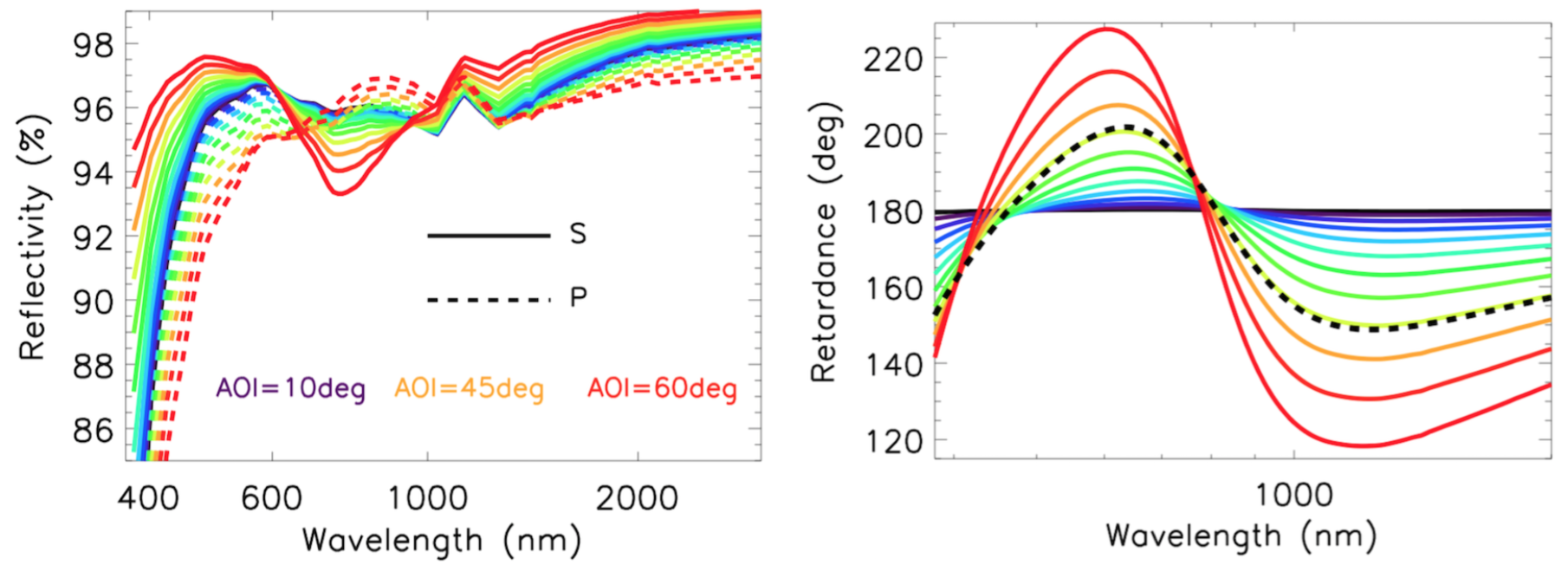}
\caption{The Zemax computed coating properties for the new enhanced protected silver coating formula as functions of incidence angle.  Each color in each panel shows a different incidence angle from 0$^\circ$ to 60$^\circ$ in steps of 5$^\circ$.  Black is 0$^\circ$, blue colors are 5$^\circ$ to 15$^\circ$ AOI and the red curve shows 60$^\circ$ AOI. The left hand panel shows the S reflectivity as solid lines and the R reflectivity as dashed lines. The diattenuation is seen as the difference between the solid lines and dashed lines.  Note how the diattenuation goes to zero at two separate wavelengths, and that those zero diattenuation wavelengths are functions of incidence angle.  The right hand panel shows the retardance as a function of wavelength for each incidence angle.  The dashed black line shows 45$^\circ$ incidence.  Retardance is a strong function of incidence angle.  Also note how the retardance crosses 180$^\circ$ twice, but that the wavelengths of these zero points are a function of incidence angle.}
\label{fig:coating_performance}
\end{center}
\end{figure}

Figure \ref{fig:coating_performance} shows the reflectivity, retardance and diattenuation for this enhanced protected silver coating formula.  For this coating design, there are two wavelengths where the 180$^\circ$ phase change from reflection is exactly met, but these points are functions of incidence angle. At lower incidence angles, the 180$^\circ$ retardance points shift to longer wavelengths. For the DKIST design, not all feed mirrors share the same incidence angle, so there will be no one wavelength where the telescope Mueller matrix is free of cross-talk.  There is a strong dependence on wavelength with 20$^\circ$ retardation amplitudes seen in the visible and near infrared at 45$^\circ$ incidence angles.

In addition, there are two wavelengths where the diattenuation is zero, but these two wavelengths are also functions of incidence angle. There will also not be any one wavelength where the telescope Mueller matrix is free of induced polarization. The actual coating formula from the various vendors providing the mirrors for all telescope and instrument optics are proprietary to the manufacturers.  However, many enhanced dielectric protective coatings have at least 2 and in some cases many layers of material deposited on top of the metal. Additional adhesion layers also complicate the formula.  We rely on these simple models to represent a close approximation to the polarization behavior as functions of the relevant variables.

\section{Flat Mirrors}

\begin{wrapfigure}{r}{0.5\textwidth}
\centering
\vspace{-2mm}
\begin{equation}
\label{flat_mueller_matrix}
{\bf M}_{ij} =
 \left ( \begin{array}{rrrr}
 \frac{1}{2}(1+\frac{R_p}{R_s})  	& \frac{1}{2}(1-\frac{R_p}{R_s}) 	& 0		& 0		\\
 \frac{1}{2}(1-\frac{R_p}{R_s})  	& \frac{1}{2}(1+\frac{R_p}{R_s}) 	& 0		& 0		\\
 0 						& 0							& \sqrt{\frac{R_p}{R_s}}C_\delta	& \sqrt{\frac{R_p}{R_s}}S_\delta	\\
 0					 	& 0							& -\sqrt{\frac{R_p}{R_s}}S_\delta	& \sqrt{\frac{R_p}{R_s}}C_\delta	\\ 
 \end{array} \right ) 
\end{equation}
\vspace{-4mm}
 \end{wrapfigure}

We present in this section some simple Zemax computations with flat mirrors. These calculations are readily comparable with theory and various lab tests. Most optical ray tracing programs, including Zemax, will output reflectivity, diattenuation and retardance for a single coated surface. The theoretical calculation involves three parameters.  $R_s$ is the reflectivity in the S- plane (German: senkrecht, meaning perpendicular). $R_p$ is the reflectivity in the P- plane (German: parallele meaning parallel).  The retardance is denoted as $\delta$.  With the notation of $C_\delta$ and $S_\delta$ denoting Cosine and Sine respectively, a simple Mueller matrix for a single flat fold mirror in a collimated beam is computed via Equation \ref{flat_mueller_matrix}.

DKIST staff and others in the literature have used simple Mueller matrix formulas based on a single ray at a single fold angle to estimate Mueller matrix properties \cite{vanHarten:2009gi}. In Zemax, a flat mirror in a collimated beam represents that approximation and should reproduce the simple Mueller matrix dependencies found with simple theory.

\subsection{Converting Electric Fields to Mueller Matrices in Converging Beams}

\begin{wrapfigure}{l}{0.30\textwidth}
\begin{center}
\vspace{-6mm}
\includegraphics[width=0.98\linewidth, angle=0]{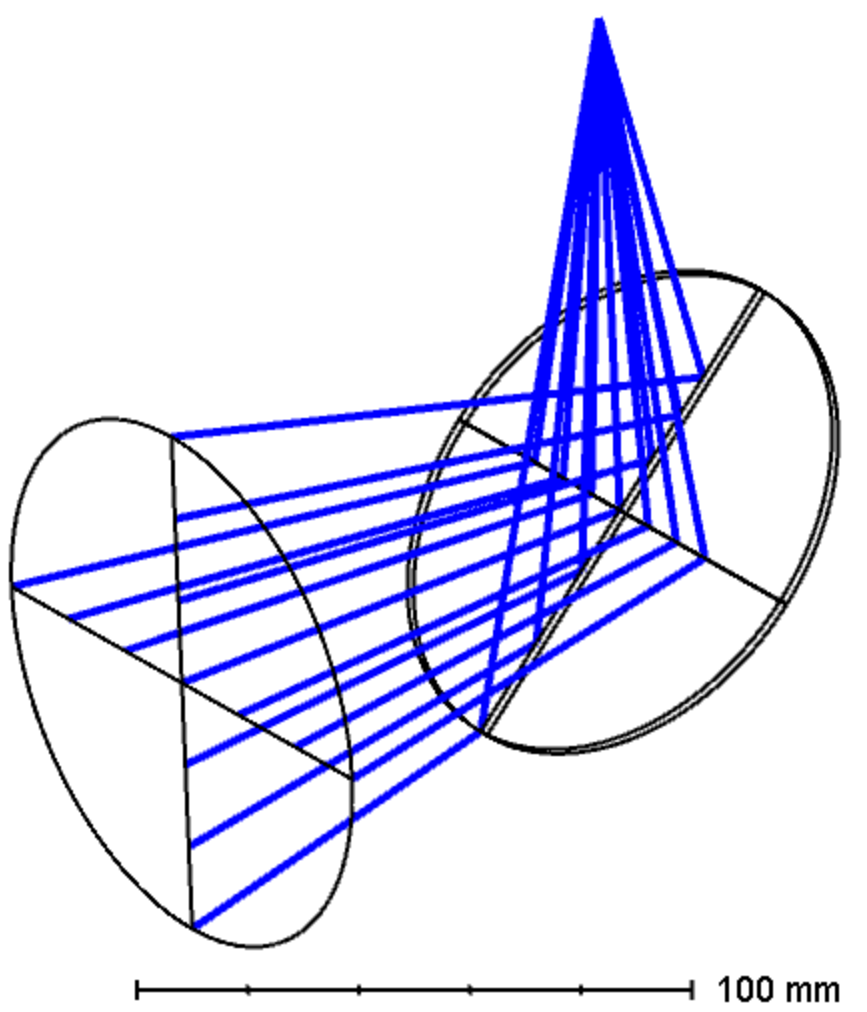}
\caption{ \label{fig:fold_f2}  The rays for the f/2 flat fold mirror at a 45$^\circ$ incidence angle. }
\end{center}
\vspace{-11mm}
 \end{wrapfigure}
A flat fold mirror model was created in Zemax.  A paraxial 200mm focal length lens was inserted at the 100mm diameter system stop. This 100mm diameter beam comes to focus after 200mm of propagation and is an f/2 system. As the entire system is paraxial, the rms spot size at the focal plane is zero within numerical accuracy.  There are no geometrical aberrations in the system.  The fold mirror was set to an incidence angle of 45$^\circ$ corresponding to a 90$^\circ$ fold.  The mirror was placed half way between the stop and focal plane, corresponding to 100mm of propagation between the stop and the focus.  Figure \ref{fig:fold_f2} shows the system layout.  For an f/2 fold, the incidence angles vary from 30.96$^\circ$ to 59.04$^\circ$. 

We demonstrate the conversion between electric field vectors and Mueller matrices by using a converging f/2 beam reflecting of a flat fold mirror at 45$^\circ$ incidence angle.  

We computed Mueller matrices for this fold mirror with one of our enhanced protected silver formulas. The $II$ and $QQ$ elements of the Mueller matrix were 0.9513.  When the Mueller matrix was normalized such that $II$=1, the $IQ$ and $QI$ terms were 0.0513, the $UV$ and $VU$ terms had an amplitude of 0.3238 with the $UU$ and $VV$ terms at 0.9447.   The coating analysis in Zemax agreed with these electric field calculations.  A fold at 45$^\circ$ incidence angle with this coating gave a retardance of 161.1$^\circ$, a diattenuation of 0.0513 and $R_s$ was 93.42\% while $R_p$ was 84.33\%.

\begin{figure}[htbp]
\begin{center}
\hbox{
\includegraphics[width=0.99\linewidth, angle=0]{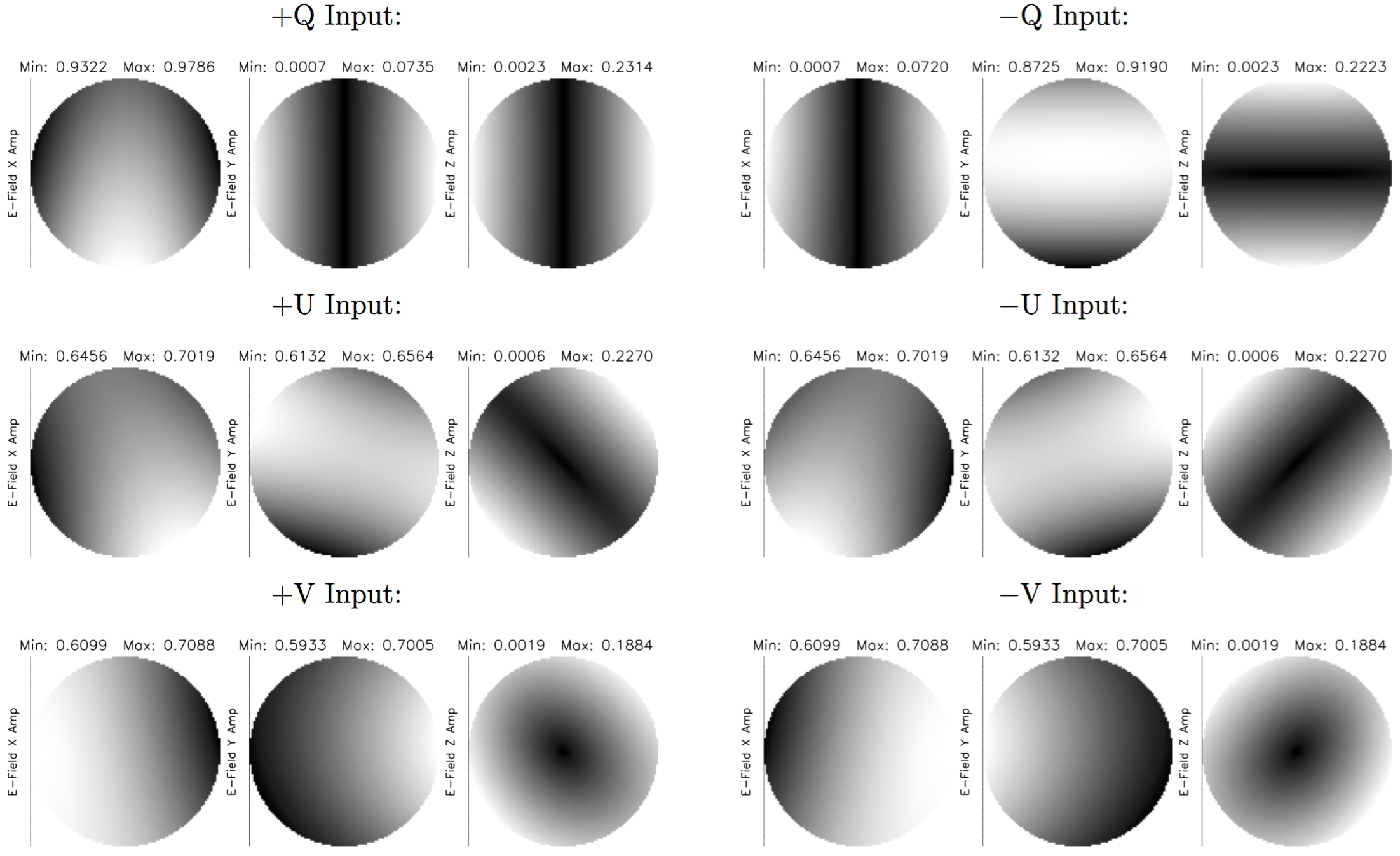} 
}
\caption{ \label{electric_field_amplitudes} Electric field component amplitudes calculated across the beam footprint for a flat fold mirror in an f/2 beam set at 45$^\circ$ AOI on the chief ray of the system shown in Figure \ref{fig:fold_f2}. The Jones matrix corresponding to the fully polarized pure Stokes inputs were specified and traced using our scripts to the focal plane of this f/2 beam.  Each panel shows a linear grey scale of the XYZ electric field amplitude from the minimum to maximum listed above each panel. As an example, the top left corner shows the amplitude of the X component of the electric field when a pure +Q input state is traced through the system linearly scaled from +0. 9322 to +0.9786. This mirror has a reflectivity of 0.9342 for the S- state at a 45$^\circ$ incidence angle but in the f/2 beam, the incidence angles vary from 30.96$^\circ$ to 59.04$^\circ$. The Y field amplitude is below 0.074.  The Z field component is below 0.23.  The top right 3 panels show the -Q input state.  This represents the -P beam with an reflectivity of 84.33\% at an AOI of 45$^\circ$. Variation is seen in the Y field amplitude from 0.87 to 0.92.  The Z field component has changed orientation but has the same amplitude as the $+Q input$. The X amplitude for $-Q$ input looks like the Y amplitude for the $+Q$ input.  Similar changes in orientation and sign are seen for the $\pm U$ and $\pm V$ input states.  A clear pattern is that the XY field amplitudes are much higher for the $\pm$Q input states while the $U$ and $V$ inputs show a mix between field amplitudes.  The $UV$ and $VU$ Mueller matrix elements are $\sim$0.33. }
\end{center}
\vspace{-8mm}
\end{figure}

Following the formula in Figure \ref{flat_mueller_matrix} and using the computed reflectivities from the Zemax coating file, we recover the same Mueller matrix computed from the electric fields.  This shows consistency between the mathematics used in the Zemax coating computations and our electric field calculations.  Note that with incidence angles varying from 30.96$^\circ$ to 59.04$^\circ$ across the footprint, there will be strong variation in the derived Mueller matrices. 

The scripts produce the electric field real and imaginary components from, which we compute the XYZ components of the electric field across the footprint.  In our case for this f/2 converging beam, the rays all converge to the focal plane and as such, are spatially overlapping.  Figure \ref{electric_field_amplitudes} shows the XYZ electric field amplitudes when all pure Jones vectors are input (representing the 6 fully polarized pure Stokes inputs). 

Zemax propagates rays in the Jones formalism by a specified optical path length along a computed propagation direction. When the electric field distribution is computed, the xyz coordinates represent that of the global xyz coordinates computed for the system at that location in space. In Figure \ref{electric_field_amplitudes}, there are substantial Z components to the electric field for all non-chief rays. 

In any real polarimeter, the performance of the analyzing polarizer, beam splitter transmission functions, reflection coefficients, etc, will all be substantially different than if one simply simulates the behavior of the chief ray.  As a simple computational aide, we collimate the incoming f/2 beam with a paraxial lens.  When collimated, the Z component of the electric field went to zero.   The electric field computation we outlined above, the Mueller matrix is derived from the Stokes parameters considering only the XY components of the electric field. In most situations, the Z component of the field is small as the f/ number of the beam in the polarimeter is typically greater than 10.  A more detailed computation with the full 3D electric field distribution is possible and will be explored in future works.

The properties of the Mueller matrix change substantially with the f/ number of the beam.  When considering the normalized Mueller matrix with only the fold at 45$^\circ$, the $IQ$ and $QI$ terms were 0.0513, the $UV$ and $VU$ terms had an amplitude of 0.3238 with the $UU$ and $VV$ terms at 0.9447.   When summing over the footprint in the f/2 beam, we see a mild increase in the $IQ$ and $QI$ terms to 0.0517. The $UV$ and $VU$ terms change by about 5\% to 0.3065 amplitude. The $UU$ and $VV$ terms also change to 0.9492.  

\begin{wrapfigure}{r}{0.55\textwidth}
\begin{center}
\vspace{-8mm}
\hbox{
\includegraphics[width=0.99\linewidth, angle=0]{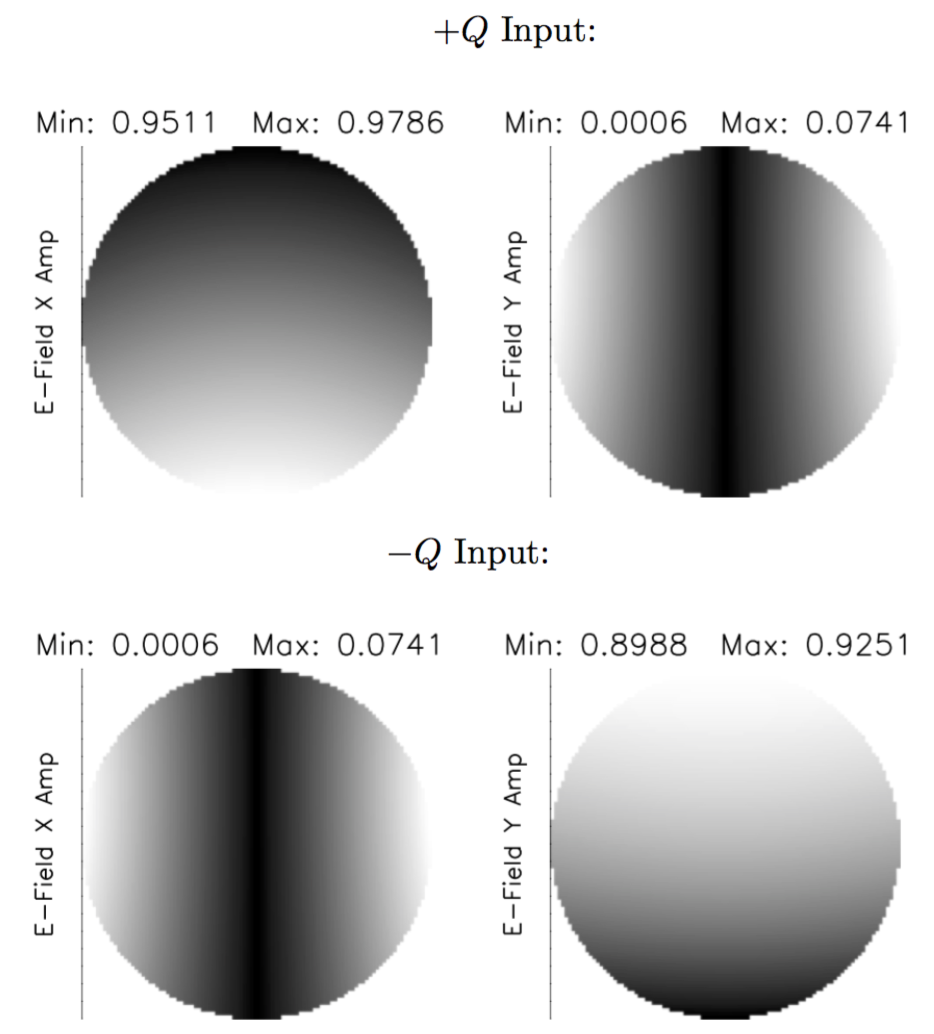}
}
\caption{ \label{fig:e_fields_collimated}  The electric field components calculated across the beam footprint similar to Figure \ref{electric_field_amplitudes} but after a paraxial collimating lens was placed near focus. With the collimated beam, the Z component of the electric field is zero.  We only show $\pm$Q input states here for comparison with Figure \ref{electric_field_amplitudes}.  The $+Q$ input state electric field component for X ranges from 0.951 to 0.979 here, whereas it was from 0.932 to 0.979 in the f/2 system of Figure \ref{electric_field_amplitudes}. }
\end{center}
\vspace{-10mm}
 \end{wrapfigure}

In addition, the $QQ$ term is no longer identical with the $II$ term. In the normalized Mueller matrix, all terms are divided by $II$ but the $QQ$ term is 0.9968 after normalization, showing some depolarization.  The diattenuation, depolarization and retardance of the system can become much more complex than the formula for a simple single fold mirror. 

We should note that the variation in electric field properties seen in Figure \ref{electric_field_amplitudes} is dominated by AOI variation across the mirror.  We have run calculations where we use a second paraxial lens near the focal plane to collimate the system.  As expected, the Z component of the electric field goes to zero.  Figure \ref{fig:e_fields_collimated} shows the amplitude of the X and Y electric field components when the converging beam is collimated near focus.  The Z component of the electric field is zero.  With the symmetries of the Z field component and the use of both the $+$ and $-$ Stokes inputs to derive the system Mueller matrix, the numerical values of the Mueller matrix are the same whether the system is collimated or converging f/2. 

\begin{figure}[htbp]
\begin{center}
\includegraphics[width=0.98\linewidth, angle=0]{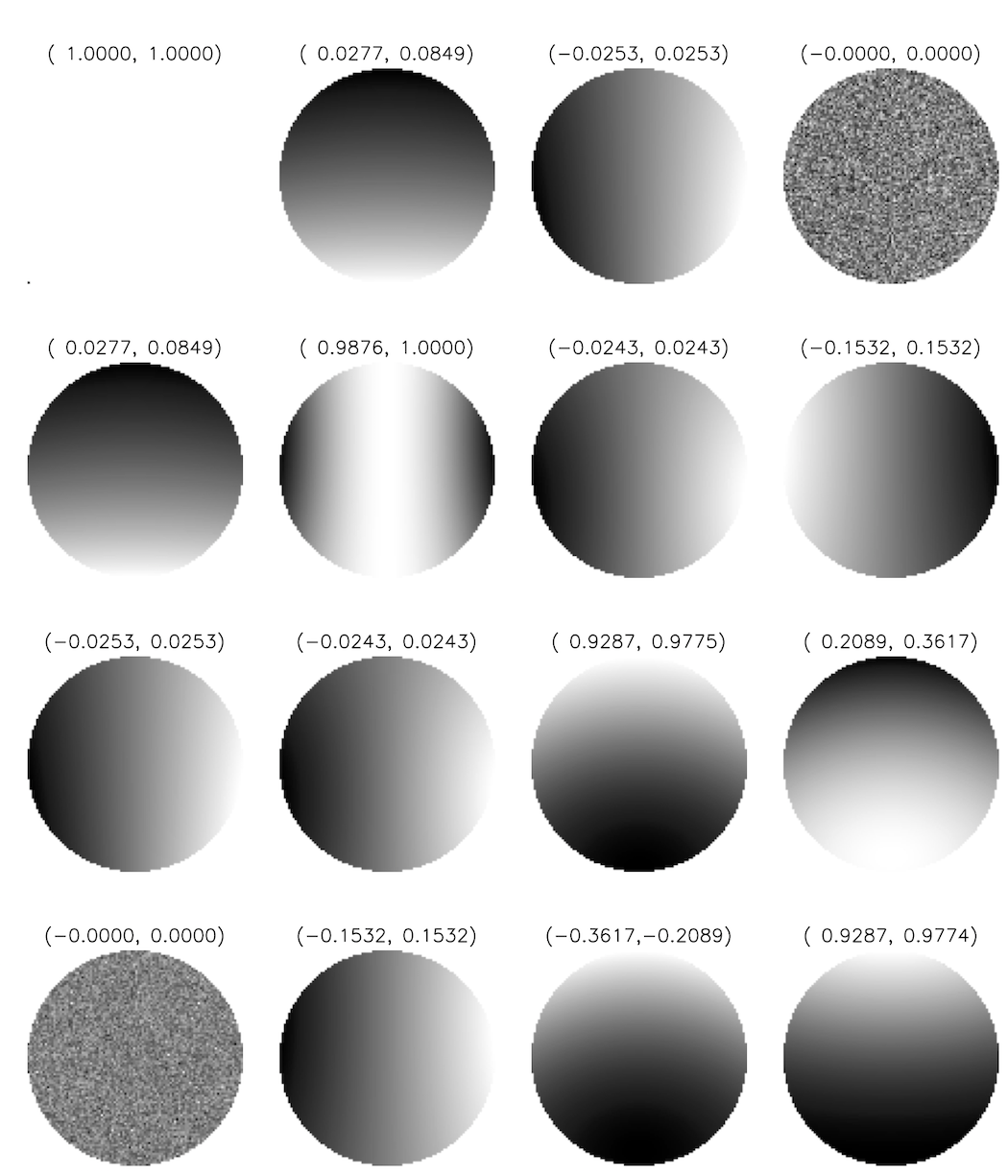}
\caption{The Mueller matrix across the beam footprint for the f/2 fold normalized by the intensity. The Stokes vectors were computed after paraxial collimation near the focal plane to set the Z component of the electric field to zero. The incidence angles vary from 30.96$^\circ$ to 59.04$^\circ$ along the extreme marginal rays reflecting off the fold mirror but the incidence angles are set to zero across the footprint on the focal plane by the collimating paraxial lens. The intensity normalization is done for each ray independently and the $II$ elements ran from 0.855 to 0.888.   Each panel shows the Mueller matrix element with the linear grey scale limits from min to max.  The $QQ$ term is always above 0.9876.  The $UV$ term has amplitudes ranging from 0.21 to 0.36. }
\label{fig:Mueller_Matrix_f2_fold}
\end{center}
\end{figure}

Figure \ref{fig:Mueller_Matrix_f2_fold} shows the derived Mueller matrix from the f/2 fold computed after a collimating paraxial lens, which sets the Z component of the electric field to zero for all rays. Note, when the system is not collimated on the optical surface evaluated,  there is an asymmetry between the top row of I to $QUV$ terms and the first column of $QUV$ to $I$ terms in the derived Mueller matrix.  As an example, this f/2 fold gives the Mueller matrix in Figure \ref{fig:Mueller_Matrix_f2_fold} when collimated but when at f/2, the $VI$ term ranges about $\pm$1\% while the $IV$ term is unchanged and is zero with numerical precision limits.

As we use only the XY components of the electric field to compute the Stokes vectors from the Jones formalism, we make an error in steeply converging beams. A real polarimeter uses an optic to analyze the beam, propagating some components of the electric field vector that also vary strongly with incidence angle and the type of optic used.  An open question to be investigated in future work is what the limits are as a function of f/ number on a real analyzer.  For the time being, we will only convert from Jones to Stokes in slow systems such as f/200 for AEOS, f/40 for HiVIS, f/26 for ViSP presented below.

\subsection{Aluminum oxide coatings and comparisons with flat mirrors reflections in the DST}

The Zemax model predictions change drastically in response to changing coating formulas. Small changes in the refractive index, absorption coefficients or thicknesses of protective layers can change the diattuenation by $>$1\% and retardance by many degrees.  If the retardation and reflectivity of the coating formula for the DKIST mirrors are not matched in detail, the system performance predictions will be inaccurate.  For off-axis, high angle of incidence systems like DKIST, the coating performance must be accurately measured with incidence angle to be modeled correctly as a formula. 

Modeling of aluminum metal coating and the aluminum oxide layer that forms over top is important for computing system reflectivity and polarization performance. Aluminum oxide has the same chemical formula as sapphire (Al$_2$O$_3$) but with amorphous (non-crystalline) structure and different birefringence properties. Various studies have been done on the polarization and reflective properties of aluminum and aluminum oxide compared to standard optical constants handbooks. \cite{vanHarten:2009gi, 2014SPIE.9147E..7CH, SocasNavarro:2011gn, Anonymous:rfWTP7rX, Harrington:2006hu}.

From the Dunn Solar Telescope (DST) staff, a formula was derived as 872nm of aluminum over 40nm of aluminum oxide by fitting the telescope polarization model (courtesy of David Elmore, private communication). Models from Socas-Navarro et al. \cite{SocasNavarro:2011gn} derive the mirror optical constants along with other properties from fits to the telescope Mueller matrix.

In this DST coating formula, the aluminum has a complex refractive index specified at many wavelengths. Certainly many other formulas are easily considered in response to other studies and with our own ellipsometer \cite{vanHarten:2009gi}.  An early DKIST study we performed also used (0.667, -5.57) and (0.7 -7.0) for the aluminum refractive index. Note that in studies by Harrington on the AEOS telescope \cite{2014SPIE.9147E..7CH, Harrington:2006hu}, the aluminum index of refraction was shown to have strong polarimetric impact. For DKIST as shown later, the aluminum oxide is only on M1 while the enhanced protected silver coatings dominate the system polarization behavior. 

In 2009, Polarization Properties of Real Aluminum Mirrors, I. Influence of the Aluminum Oxide Layer was published \cite{vanHarten:2009gi}.  They find thickness of 0.5 to 4nm of oxide.  This is in contrast to the 40nm to 50nm used in studies at the Dunn Solar Telescope \cite{SocasNavarro:2011gn}. We note that the wire grid polarizers we use have a wire thickness of $<$40nm and a pitch of 80nm such that the entire wire would oxidize and cease to function if values like 40nm were realistic.  

For our present study, we are simply demonstrating the impact of coatings on polarization performance predictions. The predicted aluminum reflectivity vs. wavelength is roughly similar to other models when using the refractive indices and coating layer thicknesses reported here. The interpolation between wavelengths in this DKIST coating file is also apparent due to the coarse wavelength sampling, but the overall behavior shows the expected reflectivity of 82\% to 87\% in the 380nm to 900nm wavelength range.

To test this work against earlier DKIST performance predictions, we use 500nm wavelength, 45$^\circ$ angle of incidence (a 90$^\circ$ fold angle), 1.625 real index for the oxide, and (0.6667, -5.5726) refractive index for the aluminum following internal DKIST reports. The resulting transmission is an exact match at 87.22\%. The reflectance for $R_s$ is 88.90\%, and $R_p$ is 85.53\% with the phase ($\delta$) of 2.53$^\circ$, which also exactly matches Equation \ref{flat_mueller_matrix}. The derived Mueller matrices follow Equation \ref{flat_mueller_matrix} to many decimal places showing that we do reproduce the theoretical equation within good limits. For an 80$^\circ$ fold at a wavelength of 500nm, the oxidized aluminum gives 86.85\% transmission, $IQ$ and $QI$ terms of 1.54\%, $UU$ and $VV$ terms of 0.8918 amplitude, $UV$ and $VU$ terms of 0.4522 amplitude and follows the theoretical equation. The intensity to linear polarization and linear polarization to intensity terms are a few percent.  Properties of aluminum coated fold mirror pairs can be compared with the common {\it turret} style solar telescopes of the DST, the German Vacuum Tower Telescope \cite{Schmidt:2003tz,2005A&A...443.1047B,2005A&A...437.1159B} and are common in designs of other solar telescope optical relays \cite{Bettonvil:2011wj,SocasNavarro:2005gv,Qu:2001un}.

\begin{figure}[htbp]
\begin{center}
\includegraphics[width=0.91\linewidth, angle=0]{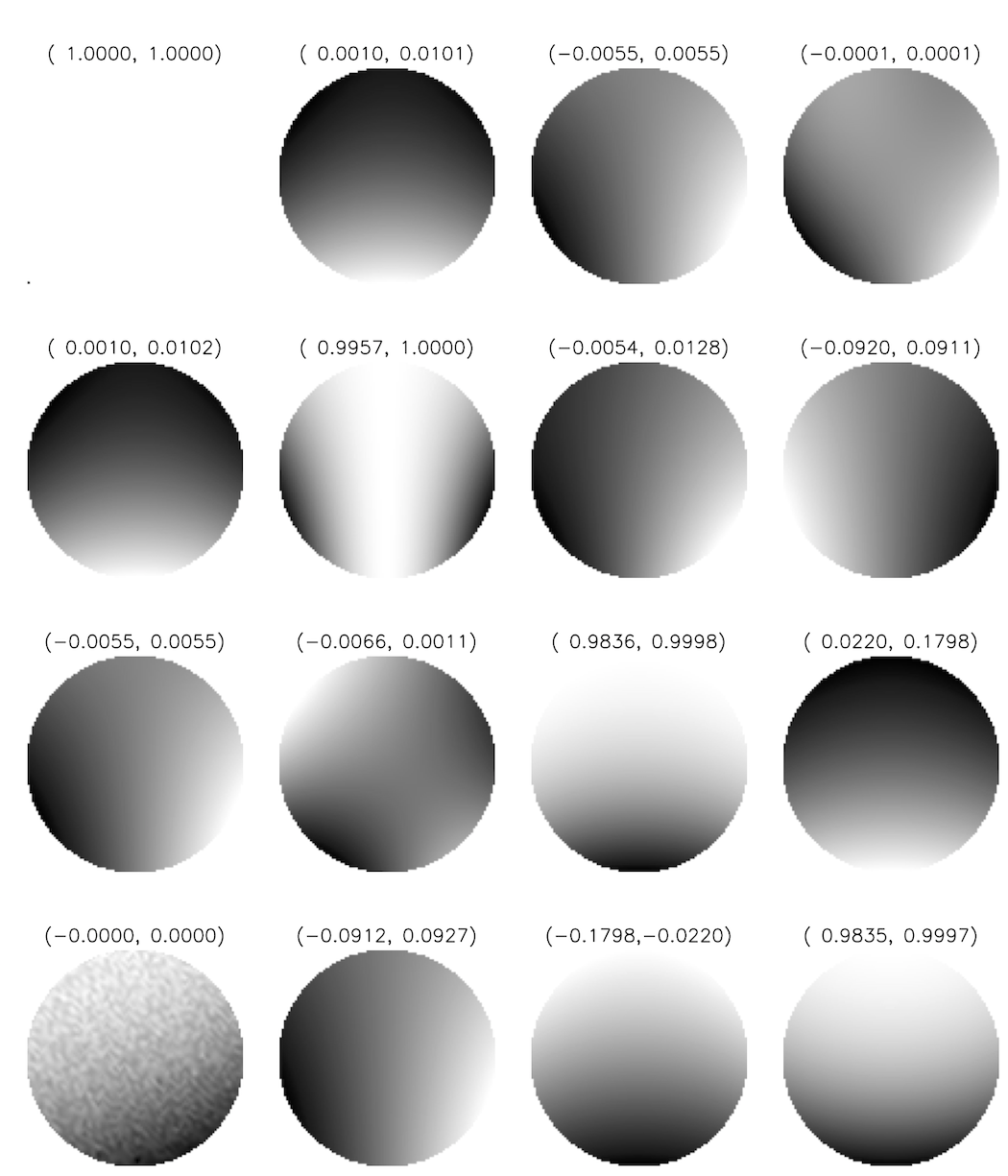}
\caption{The Mueller matrix at 400nm wavelength across the beam footprint for the f/13 Gregorian focus of DKIST for the on-axis (zero field) beam.  M1 was coated with the DST aluminum oxide formula.  M2 was coated with our enhanced protected silver formula.  The Stokes vectors were computed in the f/13 beam ignoring the Z component of the electric field vector. The incidence angles vary by $\pm$2.2$^\circ$ along the extreme marginal rays of the f/13 beam. In addition, the beam at a radius of 2.5 arc minutes is incident at a 0.33$^\circ$ angle on the field edges.  Marginal rays will thus see an asymmetry going from $\pm$1.8$^\circ$ to $\mp$2.5$^\circ$.  The intensity normalization is done for each ray independently and the $II$ elements ran from 0.720 to 0.727 accounting for the reflectivity of the oxidized aluminum and enhanced protected silver.   Each panel shows the Mueller matrix element with the linear grey scale limits from min to max.  The $QQ$ term is always above 0.9957.  The $UV$ term has amplitudes ranging from 0.022 to 0.180.  As the Z component of the electric field was ignored, there is some residual $VI$ term error of 0.0001 caused by this assumption in an f/13 beam. A separate computation with a collimating paraxial lens near Gregorian focus working f/36, paraxial f/74 reduced this to 10$^{-5}$}
\label{fig:Mueller_Matrix_GOS_Footprint}
\end{center}
\end{figure}

\section{DKIST Gregorian Focus}

\begin{wrapfigure}{l}{0.5\textwidth}
\begin{center}
\vspace{-7mm}
\includegraphics[width=0.98\linewidth, angle=0]{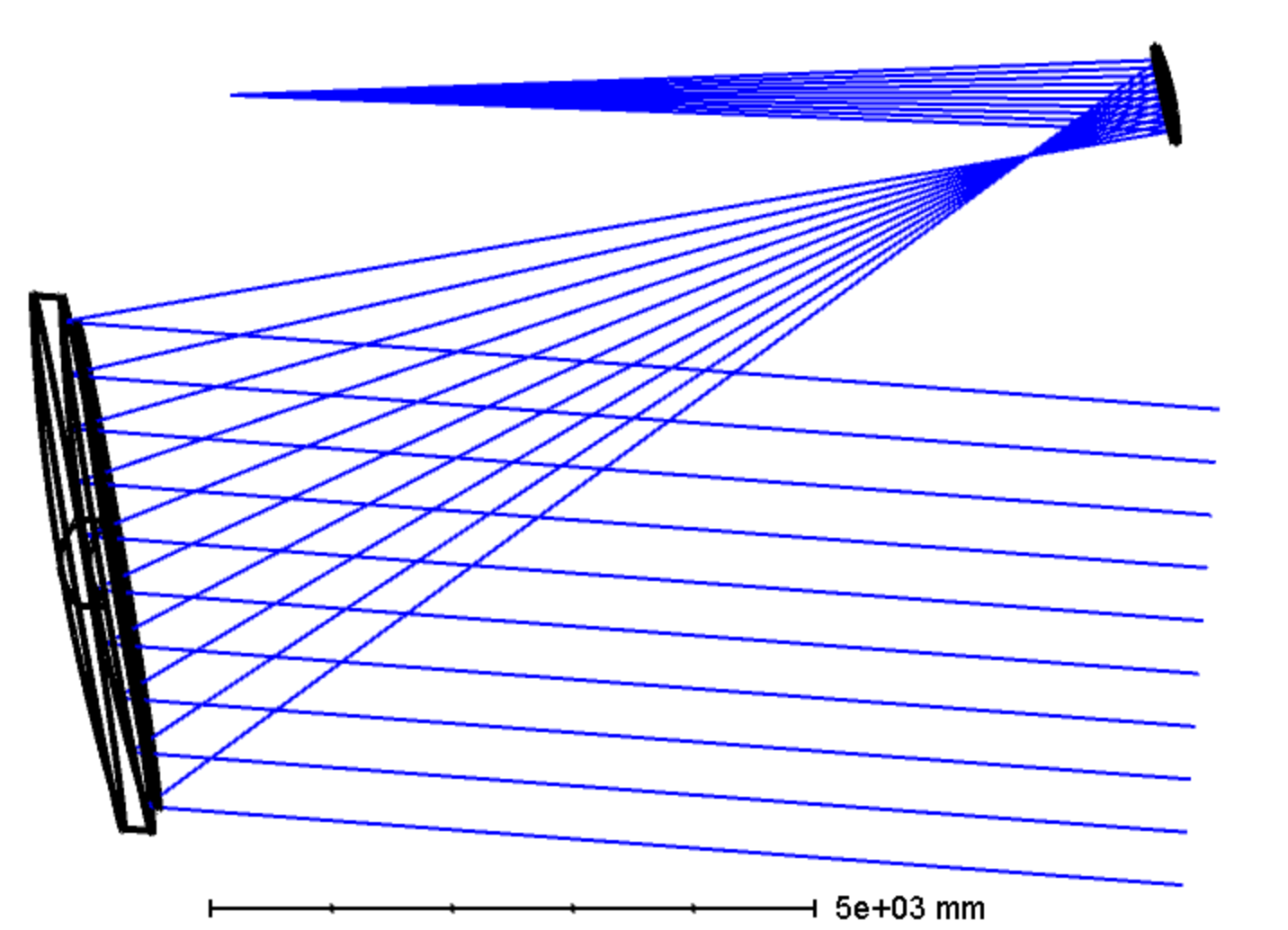}
\caption{ \label{fig:zemax_gregorian}  The on-axis rays for the beam entering the telescope and propagating to Gregorian focus. The 4.24m primary mirror with the 4.0m entrance aperture are seen creating the f/2 prime focus. Incidence angles on M1 range between 7.13$^\circ$ and 20.56$^\circ$. The incidence angles range by $\pm$2.2$^\circ$ across the f/13 beam at Gregorian focus. }
\end{center}
\vspace{-7mm}
 \end{wrapfigure}

Zemax calculations were performed to compare the baseline DKIST coating files against early reports and predictions for the Gregorian focus Mueller matrix. The oxidized aluminum formula is coated on M1 while the an enhanced protected silver formula is coated on M2.  The total intensity is around 0.87 at 500nm wavelength, which includes reflection losses by coatings (mostly from aluminum at that wavelength).  Figure \ref{fig:zemax_gregorian} shows the layout of the first two DKIST mirrors and the beam to Gregorian focus.

An internal 2002 DKIST report showed calculations and trade-offs for the polarization of prime and Gregorian focus concluding that the field dependence was well below calibration limits and was thus negligible.  This working assumption was carried forward in all DKIST documentation and is also supported here. The 2002 report was based on an f/30 Gregorian focus design, though the present DKIST design is now at f/13.  However, the report concluded that variation across the Gregorian field of view was negligible.  {\it "The off-axis elements are all well below 10$^{-5}$, and therefore no calibration would be needed even at 2.5 arc minutes away from the center of the field-of-view."}  

The properties of the current f/13 Gregorian focus are investigated below. Table \ref{table:MM_GOS} shows the Mueller matrix computed by the scripts for an f/13 beam and a slow beam (effective f/30) at 400nm wavelength. The total transmission ranges from 72.05\% to 72.76\% across the footprint. The Mueller matrix elements in the f/13 beam have some small amplitude elements ($UI$, $IU$, $QV$, $VQ$, $QU$, $UQ$) that are not present in the theoretical formula for a reflected Mueller matrix.  However, neglecting the Z electric field component influences some components of the Mueller matrix at f/13.  For the f/30 beam calculation in the lower half of Table \ref{table:MM_GOS}, these elements go to zero.  There are still some depolarization terms even when the calculations are done in a paraxially collimated system.  The $IQ$ and $QI$ terms are 0.45\%, the $QQ$ term is not 1 and the $UV$,$VU$ terms have 0.0847 amplitude.  This depolarization originates in the averaging over the aperture \cite{Noble:2011wx,2012ApOpt..51..735N,2012OExpr..20...17N, Chipman:2010tn,2005ApOpt..44.2490C,Chipman:2006iu}.  Figure \ref{fig:Mueller_Matrix_GOS_Footprint} shows the Mueller matrices varying across the footprint of the Gregorian focus beam.  Substantial variation is seen along with some asymmetries.

\begin{wraptable}{l}{0.35\textwidth}
\vspace{-4mm}
\caption{Gregorian focus Mueller matrix}
\label{table:MM_GOS}
\centering
\begin{tabular}{r r r r}
\hline\hline
1.0000 	& 0.0044 	& 0.0001 	& 0.0000  \\
0.0044 	& 0.9991 	& -0.0028 	& 0.0007  \\
0.0001	& 0.0028 	& 0.9956	& -0.0841  \\
0.0000 	& -0.0004	& 0.0842 	& 0.9946 \\
\hline
\hline
1.0000	& 0.0045	& 0.0000	& 0.0000  \\
0.0045	& 0.9991	& 0.0000	& 0.0000 	\\
0.0000	& 0.0000	& 0.9955 	&-0.0847 	\\
0.0000	& 0.0000	& 0.0848	&  0.9946	\\
\hline
\end{tabular}
The Mueller matrix computed for Gregorian focus at 400nm wavelength, f/13 above, f/30 and collimated below. Depolarization is seen $\sim$0.1\% in the diagonal as $QQ \ne 1$ and $UU \ne VV <$ 0.9964 as $Cos(Sin^{-1}(0.0847))$.
\end{wraptable}

We compute Mueller matrix variation for both the on-axis footprint and the 5 arc minute field.  As expected for an f/2 off-axis system, there are substantial asymmetries in the footprints for all field angles.  However, most of the Mueller matrix terms are symmetric and largely average to zero after reflecting off the M1-M2 mirror pair.  As concluded in a 2002 DKIST report, the Mueller matrix variation across the 5 arc minute field of view is at the 10$^{-5}$ amplitude level and is well below other DKIST calibration issues. 
 
\begin{figure}[htbp]
\begin{center}
\includegraphics[width=0.99\linewidth, angle=0]{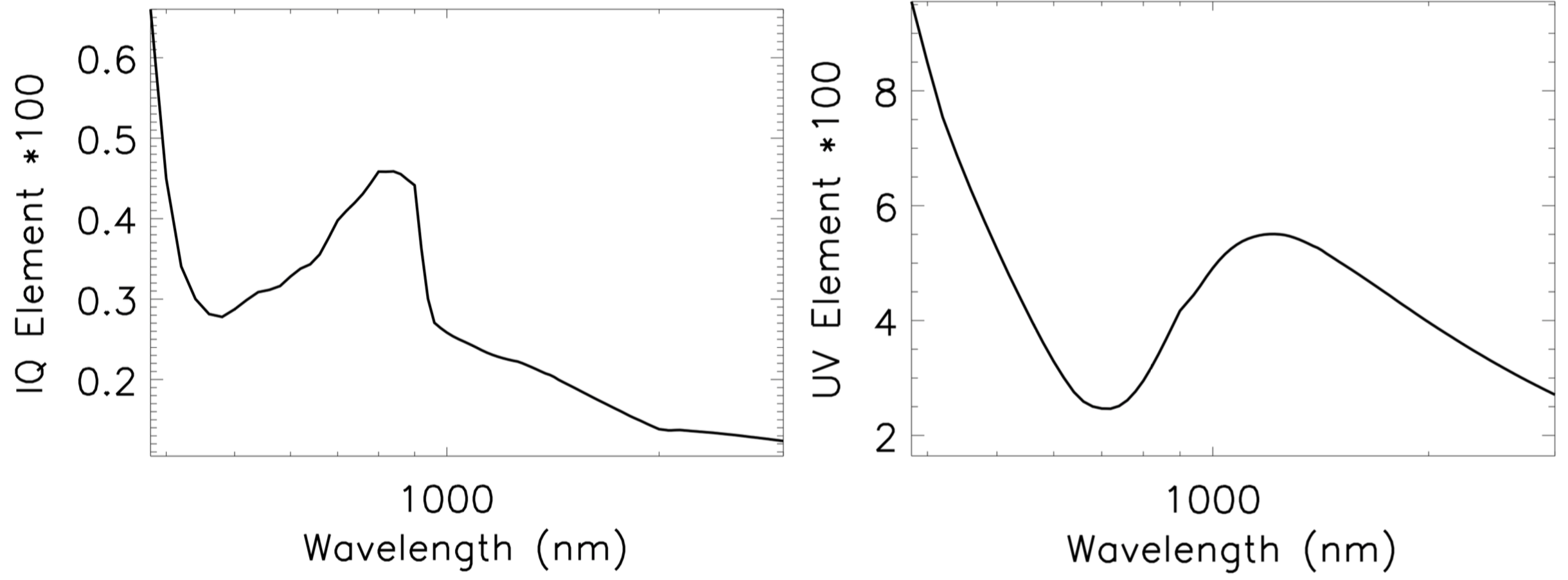}
\caption{The Zemax computed $IQ$ and $UV$ Mueller matrix elements for the on-axis beam at Gregorian focus.  The oxidized aluminum coating was used on M1 and the enhanced protected silver formula was used on M2. Aluminum is less reflective around 800nm wavelength and this creates an increase in diattenuation.  Both coatings have wavelength dependent retardance, which creates similarly complex behavior with wavelength in the $UV$ element of the Mueller matrix.}
\label{fig:gregorian_IQ_UV}
\end{center}
\end{figure}

We show in Figure \ref{fig:gregorian_IQ_UV} the expected amplitude of some Mueller matrix elements.  The $IQ$ and $QI$ terms show a higher amplitude at short wavelengths, caused by the silver, and a peak at 800nm of about 0.5\%, caused by the aluminum.  Similar effects are seen in the amplitude of the $UV$ and $VU$ elements on the right side of Figure \ref{fig:gregorian_IQ_UV}.  The Matrix element is above 0.08 at short wavelengths and falls towards the visible band but again rises in the NIR, mostly caused by the silver coating.

\section{CryoNIRSP Predictions with Pointing, Wavelength and Field}

\begin{figure}[htbp]
\begin{center}
\hbox{
\includegraphics[width=0.98\linewidth, angle=0]{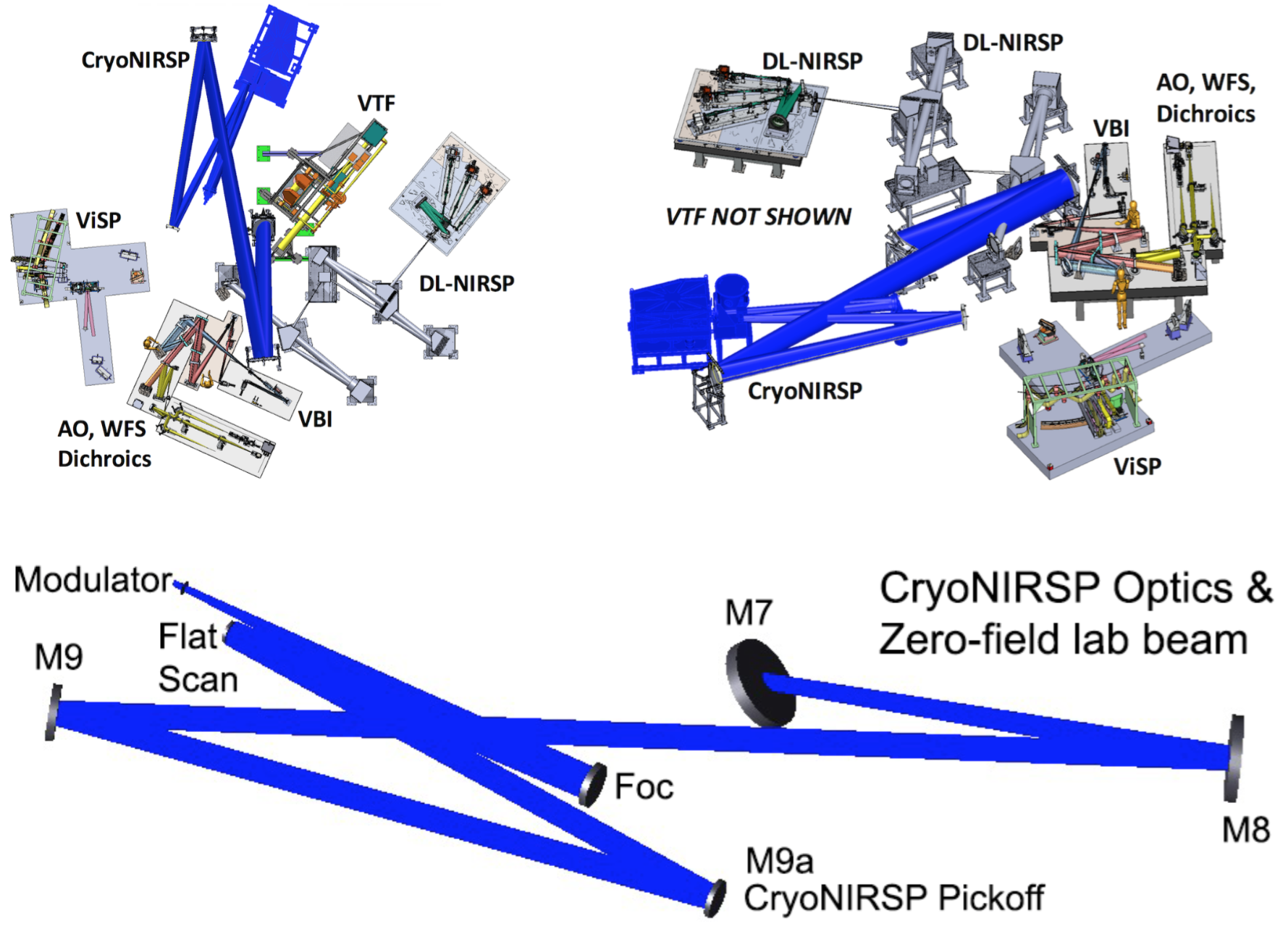}
\hspace{0mm}
}
\caption{The CryoNIRSP feed optics on the coud\'{e} floor from M7 through the modulator. The top left image shows a solid model of the coud\'{e} lab instruments with the beam and CryoNIRSP highlighted in blue.  The top right image shows the solid model from a side view again with CryoNIRSP and the beam highlighted in blue with the VTF instrument removed for clarity. The solid models have the ViSP beam removed for clarity.  The bottom panel shows the optical model with zero field of view.  M7 folds the incoming f/53 coud\'{e} beam parallel to the lab floor.  M8 is an off axis parabola that collimates the beam which is then folded by M9.  The M9a pickoff flat mirror is an enhanced protected silver fold that is inserted to feed all light to CryoNIRSP. The CryoNIRSP instrument optics include a beam splitter at 9$^\circ$ incidence angle, a scanning mirror at 4$^\circ$ incidence angle and another off axis parabola at very small angle to accomplish focusing on to the slit through the modulator. The scanning mirror can repoint the instrument against the incoming beam, effectively de-centering the beam of this slit instrument to off-axis footprints on upstream optics.  \label{cn_solid_model}}
\end{center}
\vspace{-6mm}
\end{figure}

We created polarization models for all optics feeding the CryoNIRSP instrument as functions of wavelength, telescope pointing and field of view. Figure \ref{cn_solid_model} shows top and side views of the coud\'{e} lab solid model and the optical design for the coud\'{e} lab optics between the 7$^{th}$ mirror (M7) and the CryoNIRSP modulator. We used the enhanced protected silver coating on all the DKIST optics from M2 to the CryoNIRSP modulator. As the CryoNIRSP instrument requires inserting a fold mirror, the beam path to CryoNIRSP is all-reflective.  Though the coatings for CryoNIRSP optics are not yet completed, we anticipate that the coating formulas presented here will be used on all relevant CryoNIRSP optics. The telescope azimuth, elevation and wavelength were varied in the scripts to articulate the system as in Figure \ref{fig:articulated}.

The Zemax models were computed at wavelengths of 500nm to 5000nm in steps of 500nm. The telescope azimuth and elevation was computed over the full hemisphere in steps of 2$^\circ$ to capture the complete (Az,El) behavior of the system Mueller matrix.

\begin{figure}[htbp]
\begin{center}
\hbox{
\hspace{-0.0em}
\includegraphics[width=0.98\linewidth, angle=0]{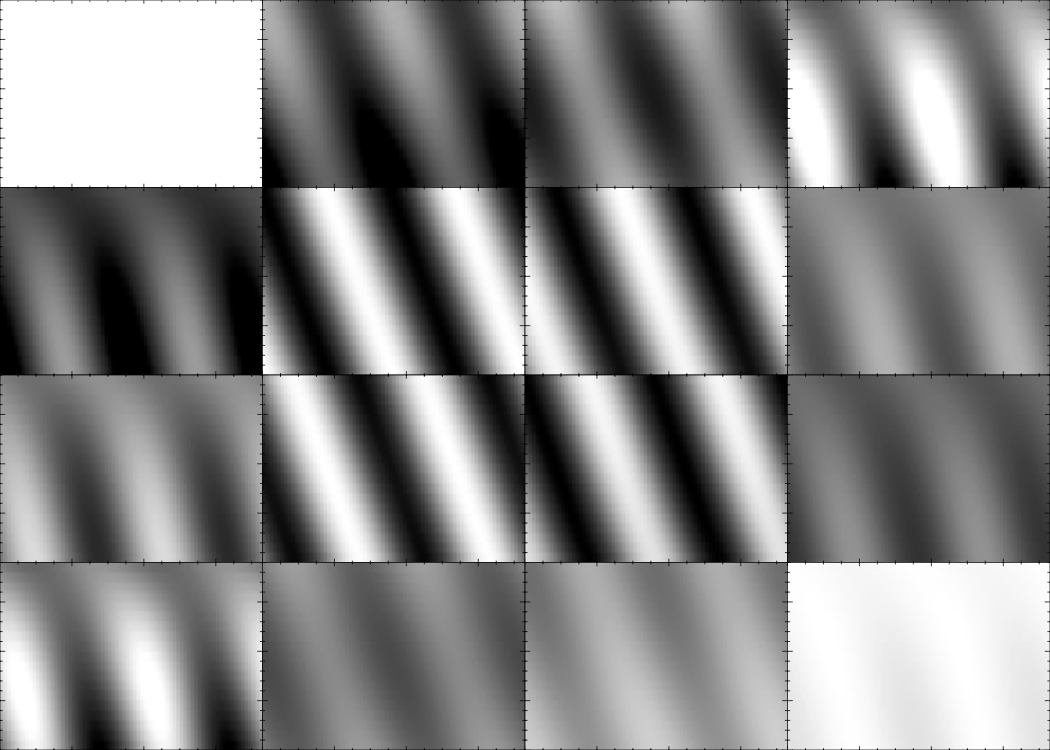}
}
\vspace{5mm}
\caption{The Zemax calculated Mueller matrix at the CryoNIRSP modulator for all telescope azimuths and elevations are shown in each panel for zero field at a wavelength of 2500nm. Each box shows a Mueller matrix element as a function of azimuth (X) and elevation (Y).  Azimuth is increasing horizontally on the X axis from 0$^\circ$ at left to 360$^\circ$ at right of each element. Elevation is increasing vertically (Y) from the horizon (0$^\circ$) at the bottom to the zenith (90$^\circ$) at the top. Thus, each box shows a Mueller matrix element over the full hemisphere (azimuth from 0$^\circ$ to 360$^\circ$ and elevation from 0$^\circ$ to 90$^\circ$) of possible optical pointings, which is beyond the actual capabilities of the telescope mechanical structure. This clearly shows the simple functional form of the Mueller matrix with azimuth and elevation. Each Mueller matrix element can be represented as Sin and Cos functions of 2*Azimuth, Elevation and 2*Elevation. This is caused by the crossing and un-crossing of the S- and P- planes of the fold mirrors.  This is also the same as shown for the AEOS telescope \cite{Harrington:2017ej} where on-sky calibrations have the same dependence and were easily fit with simple trigonometric functions. The $QUV$ to $QUV$ terms are linearly scaled to amplitudes of $\pm$1. The intensity to $QU$ terms and the $QU$ to intensity terms  ($IQ, QI, IU, UI$) are scaled to $\pm$1\%.  The $IV$ and $VI$ terms are scaled to $\pm$0.2\%. For all these $I$ to $QUV$ and $QUV$ to $I$ terms, the the $Cos(2Az)$ dependence clearly stands out with an additional dependence on elevation. These terms reach their maximum amplitudes at some (azimuth,elevation) combinations with zero polarization seen only at specific pointings.  As the coordinates are fixed with respect to the system entrance pupil for this calculation, there is a strong geometrical rotation seen in the $QU$ to $QU$ terms caused by simple coordinate transformation.  However, linear to circular cross-talk is indeed present at some telescope pointings, as seen by the variation in the $QV$ and $UV$ elements.  The $VV$ term begins to drop away from 1 at low elevations. \label{cryonirsp_mueller_map_azel}}
\end{center}
\end{figure}

The computation was done in a coordinate reference system that is tied to the entrance pupil of the optical train. This causes the definition of the $QU$ coordinate grid to rotate as seen from the perspective of a fixed XYZ frame where the Mueller matrix is computed. This means that, in addition to any circular retardance in the system, there is a purely geometrical rotation from a $QU$ reference frame at a downstream optic to a $QU$ input frame in the entrance aperture of the Zemax optical design. 

Figure \ref{cryonirsp_mueller_map_azel} shows the system Mueller matrices as functions of azimuth and elevation at the CryoNIRSP modulator for wavelength of 2500nm for the on-axis beam. All 16 elements of the Mueller matrix are shown with azimuth on the X axis and elevation on the Y axis in their own box.  Each box is linearly scaled to highlight the azimuth-elevation dependence of the individual Mueller matrix elements. Each Mueller matrix has been normalized by the transmission at each individual telescope pointing so the (0,0) element is always 1.

As the DKIST relay is a series of fold mirrors articulated in azimuth and elevation, there is a simple functional form of the Mueller matrix. Each Mueller matrix element can be represented as $Sin$ and $Cos$ functions of Azimuth, 2*Azimuth, Elevation and 2*Elevation. This is caused by the crossing and un-crossing of the incidence planes of the various fold mirrors between the azimuth and elevation rotation axes. This is also the same behavior as shown for the AEOS telescope \cite{Harrington:2017ej}.  In that paper, on-sky calibrations were used to derive the system Mueller matrix and the functional dependence was easily fit with simple trigonometric functions. 

In Figure \ref{cryonirsp_mueller_map_azel}, the $QUV$ to $QUV$ terms are linearly scaled to amplitudes of $\pm$1. The $QU$ to $QU$ terms contain both the geometric rotation from the coordinate system and also any possible circular retardance.  The intensity to $QU$ terms and the $QU$ to intensity terms ($IQ, QI, IU, UI$) are scaled to $\pm$1\% and show strong COS(2*Az) dependence along with elevation variation.  The $IV$ and $VI$ terms are scaled to $\pm$0.2\% and show similar functional dependence. These terms reach their maximum amplitudes at some (azimuth,elevation) combinations with zero amplitude seen in specific elements only at specific telescope pointings.

\begin{figure}[htbp]
\begin{center}
\hbox{
\includegraphics[width=0.99\linewidth, angle=0]{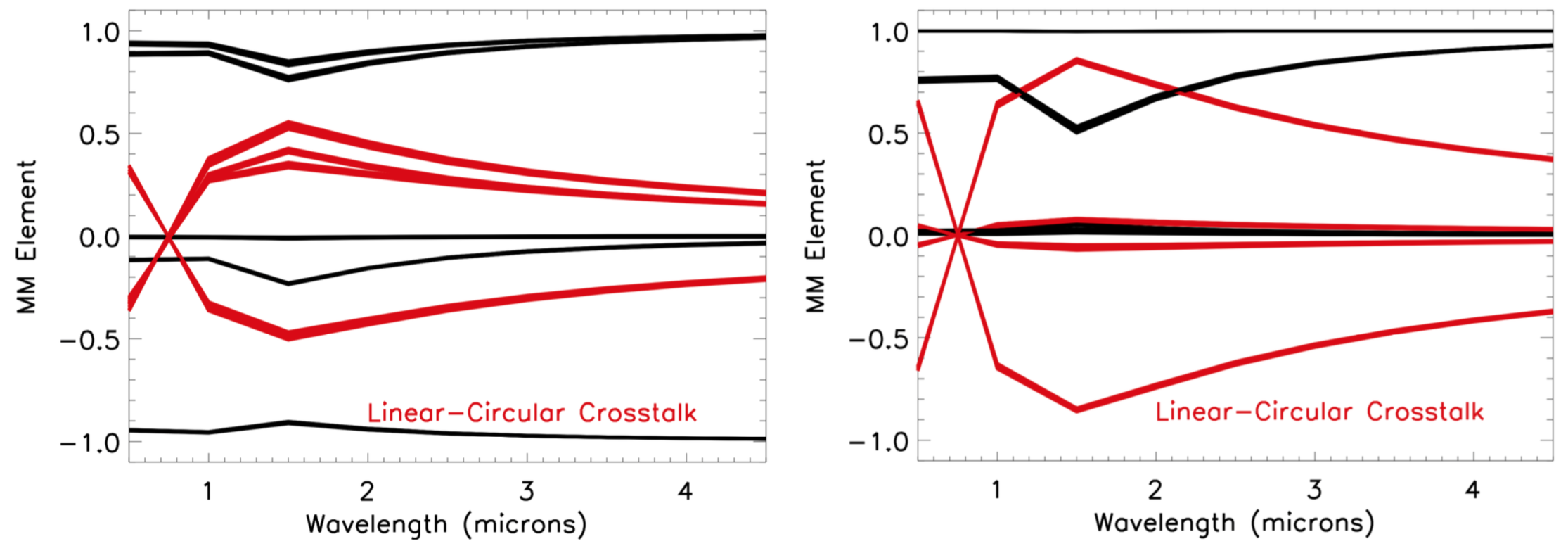}
\hspace{0.1em}
}
\caption{Mueller matrix elements $QUV$ to $QUV$ for telescope azimuth, elevation combinations of (0$^\circ$,45$^\circ$) at left and (45$^\circ$, 45$^\circ$) at right.  Red lines show the linear to circular cross-talk terms ($QUV$ to $QUV$). Some cross-talk terms reach amplitudes up to 0.5 at wavelengths around 1500nm for some pointings. Note that each panel here shows a fixed telescope pointing and represent just a single point in the (Az,El) dependence map of Figure \ref{cryonirsp_mueller_map_azel}.  \label{fig:cryonirsp_mueller} }
\end{center}
\vspace{-4mm}
\end{figure}

Given the slowly varying retardation and diattenuation of the coating formula in the 1500nm to 5000nm wavelength range from Figure \ref{fig:coating_performance}, the differences between wavelengths are mostly seen as an amplitude change of the induced polarization terms. Figure \ref{cryonirsp_mueller_map_azel} is representative of the azimuth-elevation behavior for the Mueller matrix at all near infrared wavelengths coming in to the CryoNIRSP modulator.

\vspace{-0mm}
\subsection{Field and Wavelength Variation}

Select Cryo-NIRSP Mueller matrix elements at a telescope azimuth, elevation combinations of 0$^\circ$ and 45$^\circ$ are shown in Figure \ref{fig:cryonirsp_mueller}.  The wavelength dependence is dominated by the model coating formula and the relative geometry between the groups of mirrors in the DKIST design.

\begin{figure}[htbp]
\begin{center}
\includegraphics[width=0.99\linewidth, angle=0]{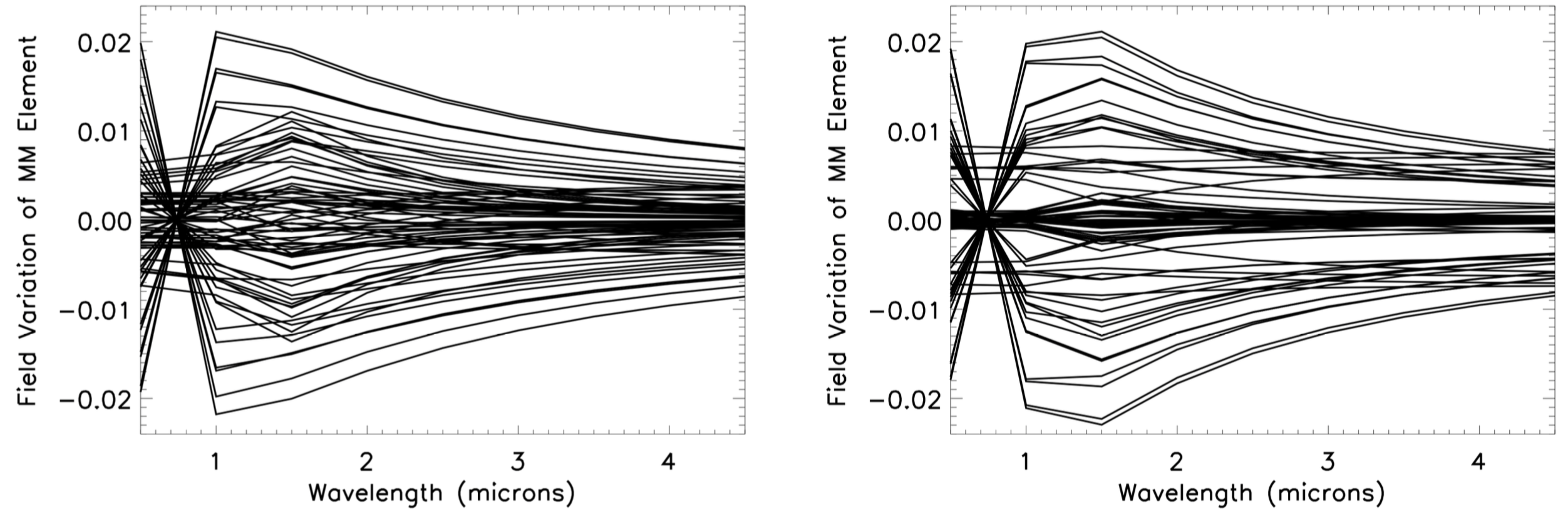}
\caption{Mueller matrix element differences between the element at field center and the element at the 5 arc minute field edge.  Each $QUV$ to $QUV$ term difference is shown for telescope azimuth, elevation combinations of (0$^\circ$,45$^\circ$) at left and (45$^\circ$, 0$^\circ$) at right. Different Mueller matrix elements have different field dependencies with wavelength. Some elements have field dependence of up to 0.02 in the value of each element. Note the zero-crossing at 850nm wavelength of some the lines as roughly the "retardance free" wavelength for the coating formula shown in Figure \ref{fig:coating_performance}. \label{fig:cryonirsp_fieldvariation} }
\end{center}
\vspace{-4mm}
\end{figure}

The predictions are generally limited by the wavelength sampling of the vendor-supplied coating data.  The coating files specified have coarse sampling in the near infrared spectral region, leading to some linear behavior with wavelength in Figure \ref{fig:cryonirsp_mueller}.  Since mirror pairs can rotate an incoming $Q$ signal in to the $UV$ cross-talk term of a subsequent mirror, Figure \ref{fig:cryonirsp_mueller} shows that some azimuth, elevation combinations have minimal $UV$ dependence regardless of the coating retardation.

If the field of view dependence is above calibration requirement amplitudes, we have to add additional variables to the calibration plans.  Figure \ref{fig:cryonirsp_fieldvariation} shows the variation from field edge to field center of the $QUV$ to $QUV$ Mueller matrix elements for a full 5 arc minute field of view. Azimuth, elevation combinations of (0$^\circ$, 45$^\circ$) are used at left and (45$^\circ$, 0$^\circ$) are used on the right.  The variation reaches amplitudes of up to 0.02 with a strong dependence on telescope pointing. The wavelength dependence generally follows the retardance formula for the coating combined with the geometric effects of one mirror rotating a linear polarization signal in to the $UV$ cross-talk axis of another mirror.  In general, this causes strong $QUV$ to $QUV$ rotations.

\section{Visible SpectroPolarimeter (ViSP)}

\begin{wrapfigure}{r}{0.60\textwidth}
\centering
\vspace{-1mm}
\includegraphics[width=0.60\textwidth]{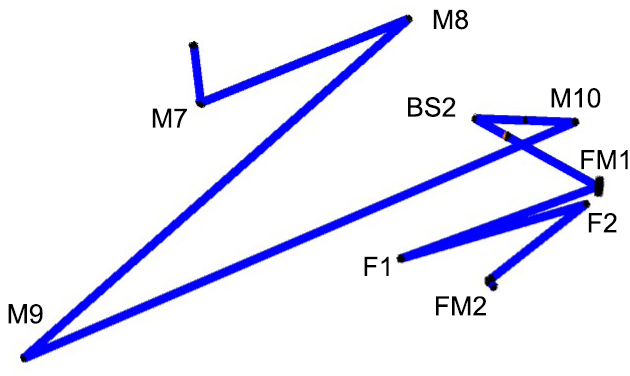}
\caption{A schematic layout of the ViSP feed optics on the coud\'{e} floor from M7 through the modulator. M7 folds the vertical beam on to the coud\'{e} floor at 45$^\circ$ incidence angle.  The DKIST optics M8 through M10 and the beam splitter BS2 feed the ViSP optics.  ViSP contains a few feed optics and fold mirrors working at a range of incidence angels.  The modulator is immediately after the final fold mirror. This fold is roughly 45$^\circ$ incidence angle. \label{fig:visp_optics} }
\end{wrapfigure}

The Visible SpectroPolarimeter (ViSP) uses several reflections and transmissions through a train of beam splitters. The beam splitters can be anti-reflection coated windows, mirrors and / or dichroic beam splitters. We chose here to show the Mueller matrix of the beam delivered to the modulator inside this instrument. The beam path on the coud\'{e} lab floor is shown for ViSP in Figure \ref{fig:visp_optics}. We use the DKIST feed optics with an all-mirror feed to ViSP with azimuth and elevation and leave analysis of the dichric beam splitters to a future paper. We can examine a case where the reflections are enhanced protected silver mirrors and the transmissions are simple uncoated window substrates.  We are in the process of modeling formulas and testing dichroic beam splitter coating samples to assess the polarization performance through many-layer coatings.

There are a few wavelengths in the visible region where the retardance of the model coating formula matches the witness sample to better than one degree. We choose three wavelengths to model the DKIST telescope optics and the ViSP instrument optics to the modulator. At 400nm wavelength, the model coating formula matches to 0.26$^\circ$.  At 600nm, the retardance matches to 0.80$^\circ$.  At 800nm wavelength the retardance matches to 0.09$^\circ$.  As of this time, ViSP has not yet selected an actual vendor to coat their mirrors so these results are approximate and can easily be re-run once we have more information about the actual coatings chosen by the team.

\begin{figure}[htbp]
\begin{center}
\hbox{
\includegraphics[width=0.98\linewidth, angle=0]{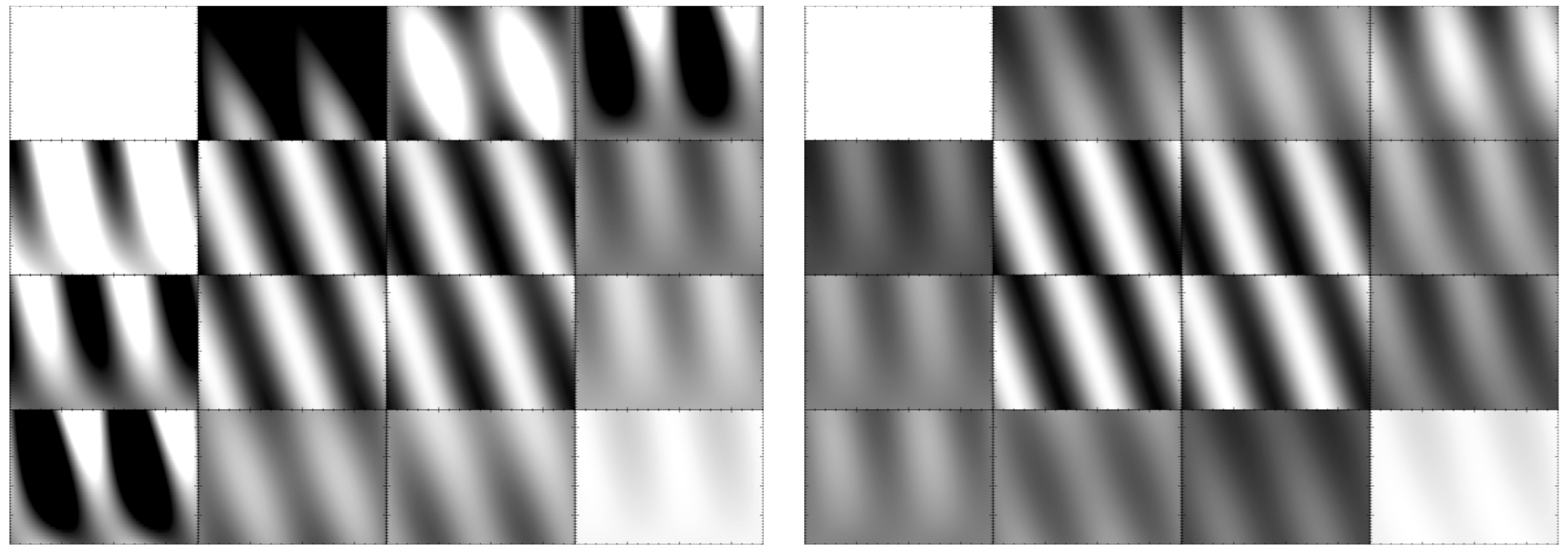}
}
\caption{The Zemax calculated Mueller matrix at the ViSP modulator for all telescope azimuths and elevations are shown in each panel for zero field. The left panel shows a wavelength of 400nm. The right panel shows a wavelength of 600nm.  Each box shows a Mueller matrix element with azimuth increasing horizontally from 0$^\circ$ at left to 360$^\circ$ at right and elevation increasing vertically from the horizon (0$^\circ$) at the bottom to the zenith (90$^\circ$) at the top. Each box shows a full hemisphere (azimuth from 0$^\circ$ to 360$^\circ$ and elevation from 0$^\circ$ to 90$^\circ$) pointing range, beyond the actual capabilities of the telescope mechanical structure. The $QUV$ to $QUV$ terms are linearly scaled to amplitudes of $\pm$1. The intensity to $QU$ terms and the $QU$ to intensity terms are scaled to $\pm$4\% for the 400nm model on the left and $\pm$2\% for the 600nm model on the right.  The $IV$ and $VI$ terms are scaled to $\pm$1.0\% for the 400nm model on the left and to $\pm$0.5\% for the 600nm model on the right.  \label{fig:visp_mueller_matrix_azel}  }
\end{center}
\end{figure}

Figure \ref{fig:visp_mueller_matrix_azel} shows the computed Mueller matrix elements while articulating the telescope Zemax design in azimuth from 0$^\circ$ to 360$^\circ$ and elevation from 0$^\circ$ to 90$^\circ$ pointing range, well beyond the actual capabilities of the telescope mechanical structure.  As expected, there is a large amplitude $QU$ to $QU$ term variation that represents the geometric rotation between the coordinates of the modulator and the coordinates of the primary mirror in addition to any circular birefringence causing $QU$ to $QU$ polarization effects.  As Zemax uses local mirror coordinates, this geometric rotation is present in all models where mirrors are articulated via Zemax coordinate breaks. 

As expected, the linear to circular polarization terms are present but are nowhere near as large an amplitude as for the AEOS telescope where we derived similar predictions \cite{Harrington:2015dl, Harrington:2017ej}. The DKIST telescope feed optics have a much more benign polarization behavior with azimuth and elevation due to the reduced incidence-angle folds.  The AEOS beam has 5 mirrors at 45$^\circ$ incidence angle whereas DKIST has [45$^\circ$, 15$^\circ$, 30$^\circ$, 45$^\circ$]. The AEOS beam has 45$^\circ$ incidence before the elevation axis, 135$^\circ$ incidence between elevation and azimuth axes, and another 45$^\circ$ to level the beam on the coud\'{e} floor.  For DKIST, these numbers are 45$^\circ$, 45$^\circ$ and 45$^\circ$.

With this coating formula at 400nm wavelength, the first 4 mirrors in the DKIST train have a diattenuation less than 5\% with ($UU$, $VV$) terms of 0.97 and ($UV$, $VU$) terms of 0.24. Though the primary and secondary mirrors have substantial incidence angles and variation across the beam, the primary $UV$ term comes from the 45$^\circ$ incidence angle of M3. The second group of mirrors are the two flat fold mirrors M5 and M6, which are at incidence angles of 15$^\circ$ and 30$^\circ$ respectively.  This group of mirrors has essentially the same linear to circular cross talk.  The diattenuation is always less than 4\% as the telescope is articulated with ($UU$, $VV$) terms of 0.96 and ($UV$, $VU$) terms of 0.27. This compares quite favorably with a configuration of three separate flat folds working at 45$^\circ$ incidence angles articulated in (azimuth, elevation) where the linear to circular terms ($UV$, $VU$) can be above 0.85 at certain pointings \cite{Harrington:2015dl}.  

For the remaining ten mirrors on the coud\'{e} lab feeding light to the ViSP modulator, the diattenuation is about 4\% with ($UU$, $VV$) terms of 0.91 and ($UV$, $VU$) terms of 0.40.  There is a 45$^\circ$ incidence angle mirror (M7), two 15$^\circ$ incidence angle mirrors (M10, BS2), a powered feed mirror at 28$^\circ$ and four other mirrors below 12$^\circ$.

\section{Summary}

We presented Zemax optical models and performance predictions for the DKIST telescope feed optics and two of the first light polarimetric instruments, the CryoNIRSP and the ViSP.  Simple flat fold mirror in a powered beam was studied to demonstrate the sensitivity to Mueller matrix elements to the f/ number of the beam. We also explored the limitations inherent in converting the Jones matrices for individual rays to the Stokes vector for the optical model using only the X and Y components of the electric field. The Mueller matrix calculations match the theoretical formula for a flat mirror based on retardance and diattenuation in agreement with theory and previous studies in the literature.  

We will be assessing models for grouping the DKIST mirrors together to predict telescope polarization as functions of field, wavelength and configuration in forthcoming publications. Beam splitter coating models are in progress and typically require 30 to 90 layers.  With these coating models, we can assess the performance of the instruments in both reflection and transmission through the beam splitters. Most of the beams fed to the first light instruments interact with the dichroic beam splitter optics 2 to 4 times.  In a future work, we will present polarization models of the other DKIST instruments accounting for the many-layer dichroic coated optics. Other ray trace programs could potentially be explored in the future. Polaris-M is an in house polarization ray tracing software developed at the University of Arizona Polarization Laboratory \cite{Yun:2010gh, Lam:2010eo, 2010SPIE.7652E..1RL, 2011ApOpt..50.2866Y, 2011ApOpt..50.2855Y}. The DKIST project has used this to model the polarimetric calibration optics at the Gregorian focus \cite{Sueoka:2016vo}. A forthcoming paper will describe the polarization performance of the calibration optics and polarization fringe predictions.

The Mueller matrix for the Gregorian focus of the DKIST primary and secondary does not have substantial field of view variation at the 10$^{-5}$ amplitudes, agreeing with previous DKIST design studies. The Mueller matrices vary substantially across any individual footprint from a single field of view due to the incidence angle variation across the highly powered primary and secondary mirrors.   These variations are substantially reduced when averaging over the footprint of the beam. 

A new model coating formula for an enhanced protected silver coating was derived to match witness sample data on coating retardance and diattenuation for the DKIST optics. With this model formula, we showed the azimuth-elevation dependence for the system Mueller matrix for CryoNIRSP and ViSP instruments as functions of field and wavelength. The Mueller matrix elements showed 2\% variations in the linear to circular polarization terms for CryoNIRSP across a 5 arc minute field.  We now have a modeling tool that allows us to compute polarization across the field of view as the telescope moves in azimuth and elevation.   With these computational tools, we can assess the quality of simple models of grouped mirrors as a way to calibrate the functional dependence of the system Mueller matrix under a wide variety of configurations and variables.

\section{Acknowledgements}

This work was supported by the DKIST project. The DKIST is managed by the National Solar Observatory (NSO), which is operated by the Association of Universities for Research in Astronomy, Inc. (AURA) under a cooperative agreement with the National Science Foundation (NSF). We thank Dr. David Elmore for his assistance, guidance and insight into the long history of work on the DKIST project. We thank Dr. Don Mickey for sharing the ZPL scripts in 2003 as part of early work on the AEOS telescope.  This work made use of the Dave Fanning and Markwardt IDL libraries. We also thank a few anonymous vendors for guidance on typical formulas for protected metal coatings.  We thank Infinite Optics for witness samples, coating formulas and assistance from knowledgable staff.

\bibliography{ms} 			
\bibliographystyle{spiebib}		

\end{document}